\documentclass{article}
\usepackage[utf8]{inputenc}

\usepackage{geometry}
\usepackage{setspace}
\usepackage{amsmath}
\usepackage{amssymb}
\usepackage{graphicx}
\usepackage{booktabs}
\usepackage{dcolumn}
\usepackage{natbib}    
\usepackage{subcaption}
\usepackage{xcolor} 

\usepackage{url}
\usepackage[colorlinks=true,citecolor=blue,linkcolor=blue,urlcolor=blue]{hyperref}
\usepackage{authblk}
\usepackage{footnote}
\usepackage{graphicx}
\usepackage{tikz}
\usepackage{bm}
\usepackage[bottom]{footmisc}

\usepackage{algorithm}
\usepackage{algpseudocode}

\title{Inferring Soil Drydown Behaviour with Adaptive Bayesian Online Changepoint Analysis}

\author[1]{Mengyi Gong\thanks{Corresponding address: School of Mathematical Sciences, Lancaster University, Lancaster, U.K., LA1 4YF, E-mail: m.gong1@lancaster.ac.uk}}
\author[1]{Christopher Nemeth}
\author[1]{Rebecca Killick}
\author[2]{Peter Strauss}
\author[3]{John Quinton}
\affil[1]{School of Mathematical Sciences, Lancaster University, Lancaster, U.K.} 
\affil[2]{Institute for Land and Water Management Research, Petzenkirchen, Austria}
\affil[3]{Lancaster Environment Centre, Lancaster University, Lancaster, U.K.}

\date{}

\begin{document}

\maketitle

\abstract{Continuous soil-moisture measurements provide a direct lens on subsurface hydrological processes, notably the post-rainfall “drydown” phase. Because these records consist of distinct, segment-specific behaviours whose forms and scales vary over time, realistic inference demands a model that captures piecewise dynamics while accommodating parameters that are unknown a priori. Building on Bayesian Online Changepoint Detection (BOCPD), we introduce two complementary extensions: a particle-filter variant that substitutes exact marginalisation with sequential Monte Carlo to enable real-time inference when critical parameters cannot be integrated out analytically, and an online-gradient variant that embeds stochastic gradient updates within BOCPD to learn application-relevant parameters on the fly without prohibitive computational cost. After validating both algorithms on synthetic data that replicate the temporal structure of field observations—detailing hyperparameter choices, priors, and cost-saving strategies—we apply them to soil-moisture series from experimental sites in Austria and the United States, quantifying site-specific drydown rates and demonstrating the advantages of our adaptive framework over static models.}

\vspace{0.5cm}
\textbf{Keywords:} structural change, segmentation, particle filter, online gradient descent, soil moisture drydown modelling \\

\section{Introduction} \label{sec:Introduction}

Healthy soil plays a critical role in sustaining biodiversity, maintaining food production, and water resources, and mitigating climate change through soil organic carbon sequestration \citep{SoilHealth, SoilReport}. Soil water content, or soil moisture content, is an important component of soil health. It is crucial to the supply of water to plants and is therefore fundamental to agricultural production. It is also a key component in the hydrological cycle, regulating the recharge of groundwater and the flow of water to surface water bodies - both are critical to ecosystem function \citep{SoilDrydown}. The last few decades have seen a growth in research on soil moisture dynamics \citep{SMspatiotemporal}, as more data from underground sensors and satellites have become available. Soil moisture drydown modelling is one area where soil scientists describe soil water loss via drainage, runoff and evapotranspiration. Drydown curves are identified from time series data, and modelled to obtain information on how moisture content changes \citep{SoilDrydown, SDdiff}.  

\begin{figure}[!htb]
\begin{center}
\includegraphics[width=5.6in]{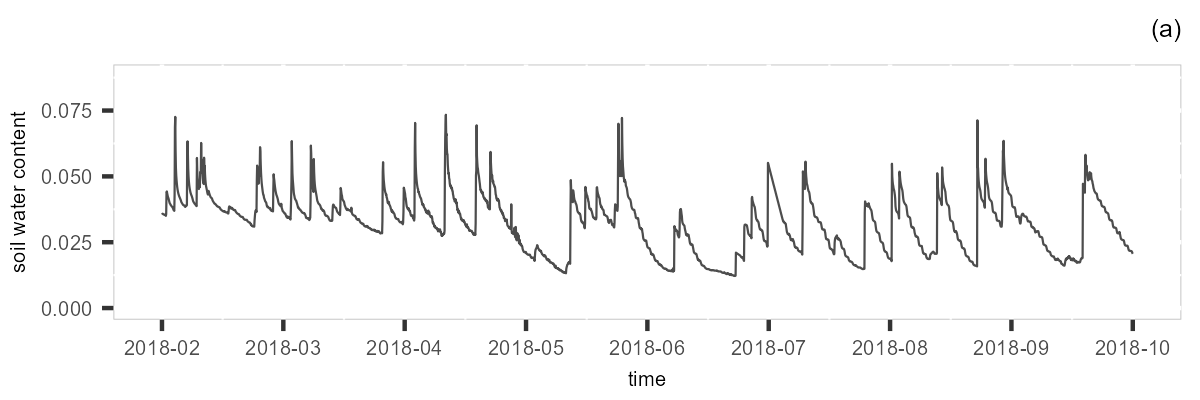}
\includegraphics[width=5.6in]{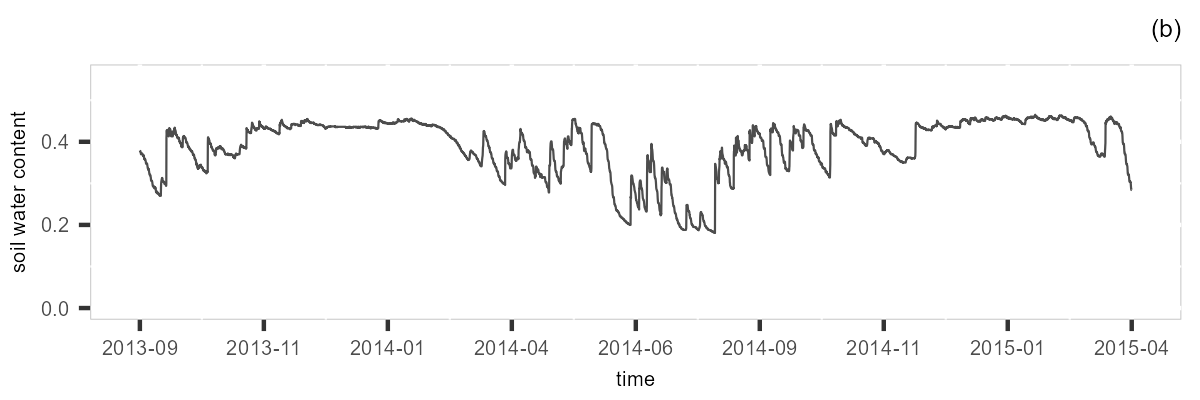}
\captionsetup{labelfont=bf, font=small}
\caption{Example of soil moisture (measured in volumetric water content) time series from (a) field site TALL of NEON in the U.S. and (b) field site 01 of HOAL in Austria.}
\label{fig:ExampleSM}
\end{center}
\end{figure}

Conventional soil drydown modelling requires manually identifying the drydown curves from soil moisture time series data, potentially with the help of additional information such as precipitation. The soil moisture drydown curves are then modelled as an exponential decay process as 
\begin{equation}
    y(t) = \alpha_{0} + \alpha_{1} \exp \left(-\frac{t}{\omega} \right) \; ,
    \label{eqn:Drydown}
\end{equation}
where $\alpha_{0}$, $\alpha_{1}$ and $\omega$ are model parameters \citep{SoilDrydown, SDdiff}. Among the three parameters, $\omega$ is of greatest interest to soil scientists, as it characterises the rate of soil moisture decay and reflects the characteristics of the soil. A median or mean of the estimated $\omega$ is often reported in the literature \citep{SDdiff, SDglobal}. To make the modelling process more automatic and to make full use of the time series data, \cite{SoilCpt} developed a changepoint-based method, where the time point before the drydown begins is considered as a changepoint. The changepoints are identified using the penalised exact linear time, or PELT, method \citep{PELT}, along with the parameters of the exponential decay model fitted to each segment. The method works well when the soil moisture time series display mostly exponential decay patterns, such as the time series in the top panel of Figure \ref{fig:ExampleSM}, a soil moisture time series from field site TALL in Alabama, U.S.A, from the National Ecological Observatory Network, or NEON, \citep{NEONsoil}. However, this is not always the case. Sometimes, due to complex precipitation and evapotranspiration patterns, such as high saturation levels of the soil or frozen soil, the soil moisture time series may display a distinctively different pattern from a typical drydown curve. For example,  during a very wet/cold period, the soil moisture content can remain high and fluctuate around this level. This can be seen in the bottom panel of Figure \ref{fig:ExampleSM}. This presents an example of the soil moisture time series from the Hydrological Open Air Laboratory, or HOAL, \citep{HOAL}, Petzenkirchen, Austria, where there is a clear difference in the drying patterns during the summer and winter seasons. Using the exponential decay model alone to describe different types of segments in the time series becomes inappropriate, as some segments are simply not governed by an exponential decay. Therefore, it would be helpful to develop a method that not only segments the data but also choose a suitable model for a given segment. Furthermore, misspecification of a segment by an exponential decay model can affect estimation of surrounding changepoints. 

Various changepoint detection methods for estimating multiple changepoints have been developed in literature, for example, frequentist methods seeded binary segmentation \citep{SeededBS} PELT \citep{PELT} and functional pruning optimal partitioning \citep{FPOP}, Bayesian methods based on Dirichlet process \citep{DirichletCpt1}, spike-and-slab priors \citep{SpikeSlab}, and hidden Markov models \citep{onlineHMM, DynamicCPT, UncertaintyCpt}. However, most of these methods do not apply to the type of changepoint detection problem in the soil moisture modelling easily, because the extensions required to incorporate multiple candidate models are non-trivial. One method that provides a basic modelling framework for the changepoint problem in this study is the Bayesian online changepoint detection method (BOCPD) proposed in \cite{BOCPD}. The observations in \cite{BOCPD} are assumed to follow a piecewise regression model, where each segment is described using a polynomial regression model of different orders. The observed process is governed by a hidden process, where the hidden state is defined to be the time of the most recent changepoint before $t$, and transition of the hidden state is assumed to be Markov. The marginal likelihood (after marginalising over the model parameters) of the segments are then used to make inference on the the changepoint locations and the polynomial orders. A related approach to Bayesian changepoint detection was proposed in \cite{BOCPD2007} where the inference was made on the run-length of each segments instead of the changepoints, and an extension of the BOCPD method was proposed in \cite{BOCPD2}, where the dependence between two adjacent segments is also considered. 

The BOCPD method can conceptually be extended to models of different types; instead of identifying the appropriate order of the regression model, we identify the type of model that is most suitable for a particular segment. One major problem of this extension is the evaluation of the marginalised likelihood. As we move from polynomial regression to a wider range of models, chances are the unknown parameter in some models will not have a conjugate prior distribution, and hence there will not be a closed-form solution to the marginal likelihood. \cite{BOCPD} suggested the use of numerical integration in these situations, but practical experiments have shown that numerical integration may not converge. In addition, it may be of interest to estimate some key parameters, such as the exponential decay parameter in the soil moisture drydown model. 

There are several approaches to address the unknown parameter problem in literature. For example, static parameters (i.e., the parameters that do not change from segment to segment), can be estimated using particle Markov chain Monte Carlo (MCMC) \citep{particleMCMC}, online EM algorithm \citep{onlineEMCPT} or sequential Monte Carlo (SMC) \citep{UncertaintyCpt}, which is sometimes referred to as a particle filter (PF). For parameters that change with segments, \cite{DynamicCPT} augments the unknown parameters to the hidden state and estimates the entire state vector via sequential Monte Carlo. Whilst these methods are generic, they do not apply to the problem considered in this manuscript straightforwardly. For example, the method in \cite{DynamicCPT} handles the segment-specific parameters nicely, but extending the method to allow for more than one type of model class is non-trivial. Depending on the model, the hidden states will take different forms, and sampling the hidden states is going to be challenging.

Despite the challenges, the literature provides insight into tackling the problem. The implicit posterior density suggests the use of Monte Carlo-based method for approximating the likelihood or gradient-based methods for parameter learning. The sequential nature of the BOCPD algorithm suggests that one can take advantage of this feature when developing the extension, which can potentially reduce the computational cost. General sequential methods have been developed over the years, such as the online EM method in \cite{onlineEM, onlineSEM} for inference of unknown parameters with closed-form sufficient statistics, the sequential Monte Carlo \citep{TutorialPF, ReviewPF} approach and online gradient approach \citep{StochasticGradient, SGDtricks} for more complex parameter estimation problems when the posteriors and marginals cannot be acquired easily. The latter two approaches are large families of methods with versatile features to suit a wide range of problems. The hidden state may be augmented with the unknown parameters or other auxiliary variables \citep{ParticleLearning, AugmentPF} before applying the particle filters to approximate the joint posteriors. 

Motivated by segmenting soil moisture time series and the literature, we propose an extension of the BOCPD method with sequential parameter update steps to segment a time series using candidate models from a pre-determined set of classes, as well as estimate the unknown model parameters that cannot be marginalised explicitly. Specifically, inference of the segment-specific parameter is carried out during the forward filtering recursions of the BOCPD algorithm. Two different approaches are considered: (1) augmenting the hidden state with the unknown parameters and approximating its posterior density using sequential Monte Carlo, or particle filter, (2) using online gradient descent to learn the unknown parameters. Both methods integrate nicely into the BOCPD algorithm due to their sequential nature. Whilst the development of the extension is motivated by the application problem in soil science, the method it self is generic and can potentially be applied to a variety of changepoint detection problems. 

The implementation details of the proposed methods are investigated and discussed, which include the choices of prior parameters and reducing computational cost when the length of the time series grows. The proposed methods are then illustrated on simulated time series, with their performance in changepoint detection and parameter inference compared. Finally, the proposed extensions are applied to the soil moisture time series from selected field sites from the National Ecological Observatory Network (NEON) in the United States (\url{https://data.neonscience.org/}) and the Hydrological Open Air Laboratory (HOAL) in Petzenkirchen, Lower Austria (\url{https://hoal.hydrology.at/the-hoal}). These field sites experience prolonged rainy periods or frozen periods, in addition to the usual drydown period during the warmer and drier seasons. Two candidate models are considered for the modelling, which are an exponential decay model (i.e., the soil moisture drydown model) for describing the dynamics in the time series during the typical drydown periods, and a mean model for describing the time series during the saturated periods or frozen periods. 

The remainder of the paper consists of five sections. Section \ref{sec:Review} provides a brief introduction to the BOCPD algorithm and the challenges to changepoint detection where there are difficult unknown parameters in some of the candidate models. Section \ref{sec:Method} describes the two extensions to the BOCPD algorithm proposed in this manuscript. Section \ref{sec:Practical} discusses the implementation details of the extensions and illustrates the methods on simulated time series data. Section \ref{sec:Application} applies the proposed extensions to the real soil moisture time series data from NEON and HOAL. Finally, section \ref{sec:Discussion} summarises the proposed method and discusses potential future improvements to the method.

\section{The Bayesian changepoint detection framework} \label{sec:Review}

In this section, we provide a brief introduction to the Bayesian changepoint detection method from \cite{BOCPD}, along with a discussion of the challenges in estimating segment specific parameters.

First we present the general changepoint model for multiple changepoints. Denote the observation at time $t$ as $y_{t}$ and the segment from time $s$ to $r$ as $\bm{y}_{s:r} = (y_{s}, y_{s+1}, \cdots y_{r})$. Assume that there are $k$ changepoints in the time series $\bm{y}_{1:n}$, denoted as $0 = \tau_{0} < \tau_{1} < \cdots < \tau_{k} < \tau_{k+1} = n$. They separate the time series into $k+1$ segments. Assume the segment between two adjacent changepoints, $\tau_{j}$ and $\tau_{j+1}$, follows the model $f_{j}(y_{t} | \bm{\eta}_{j})$ with parameter $\bm{\eta}_{j}$, $j = 1, \cdots, k+1$. Then the changepoint model can be written as 
\begin{equation}
    y_{t} = 
    \begin{cases} f_{1}(y_{t} | \bm{\eta}_{1}) + \epsilon_{t}, \; \; \tau_{0} < t \leq \tau_{1} \\ 
    f_{2}(y_{t} | \bm{\eta}_{2}) + \epsilon_{t}, \; \; \tau_{1} < t \leq \tau_{2} \\
    \; \; \vdots \\
    f_{k+1}(y_{t} | \bm{\eta}_{k+1}) + \epsilon_{n}, \; \; \tau_{k} < t \leq \tau_{k+1} \\
    \end{cases} \, .
    \label{eqn:cptProblem}
\end{equation}
Typically, the models $f_{j}(y_{t} | \bm{\eta}_{j})$ are assumed to share the same form, e.g., a linear trend model, $y_{t} = \alpha_{j} + \beta_{j}(t - \tau_{j}) + \epsilon_{t}$, and the only component that changes from segment to segment is the parameter vector $\bm{\eta}_{j} = (\alpha_{j}, \beta_{j})$. However, sometimes the segments display patterns that would require the use of more than one types of model, e.g., a linear trend model and a exponential decay model, $y_{t} = \alpha_{j} + \beta_{j}\exp(\theta_{j} (t - \tau_{j})) + \epsilon_{t}$. For the exponential decay model, the parameter vector becomes $\bm{\eta}_{j} = (\alpha_{j}, \beta_{j}, \theta_{j})$. In this case, for each segment $j$, $f_{j}( \cdot | \bm{\eta}_{j})$ can be either one of the two models, and $\bm{\eta}_{j}$ will have different contents depending on the type of model $f_{j}( \cdot | \bm{\eta}_{j})$ takes. The goal of changepoint detection is to infer the number and the locations of the changepoints. One may also be interested in inferring the type of model that is most appropriate for a segment and the corresponding model parameters.

\subsection{Bayesian inference for multiple changepoints} \label{sec:ReviewBOCPD}

In the BOCPD method of \cite{BOCPD}, the time series is assumed to follow a piecewise regression model, and the models for each segment are polynomial regression of different orders. The locations of the changepoints are modelled via a Markov process, where the transition probability, that is the probability of moving from the last changepoint $s$ to the next changepoint at $t$, is defined through a run-length distribution $g(t-s)$. Define a hidden process, $\bm{C}_{1:n}$, as the sequence of most recent changepoint prior to current time. That is, $C_{t} = s$ means the most recent changepoint before $t$ is at $s$. Then the inference of the changepoint locations can be transformed into the inference of the hidden state $C_{t}$, for $t=1, \cdots, n$. Specifically, the transition probability from $C_{t}$ to $C_{t+1}$ is 
\begin{equation}
\mathrm{Pr}(C_{t+1} = \tau | C_{t} = s) = 
\begin{cases} \frac{1 - G(t-s)}{1 - G(t-s-1)} , \;\;\;  \tau =s   \\
 \frac{G(t-s) - G(t-s-1)}{1 - G(t-s-1)} , \;\;\; \tau=t \end{cases} , 
\label{eqn:transition}
\end{equation} 
where $G(t-s) = \sum_{l=1}^{t-s} g(l)$ is the probability of having a run length up to $t-s$. A common choice of run length distribution is the geometric distribution, which results in a constant transition probability. 

Denote the order of the polynomial regression as $m$, $m=1, \cdots, q$, and denote $M_{t}$ as the choice of order of the polynomial regression model for observation $y_{t}$. Given $C_{t} = s$ and $M_{t} = m$, the marginal likelihood of the segment $\bm{y}_{(s+1):t}$ can be evaluated by integrating over the unknown parameter $\bm{\eta}$ give prior distribution $\pi(\bm{\eta})$, as
\begin{equation}
L(s, t, m) = \int f(\bm{y}_{(s+1):t} | C_{t}=s, M_{t}=m, \bm{\eta}) \pi(\bm{\eta}) \mathrm{d} \bm{\eta} \; .
\label{eqn:intLike}
\end{equation}
The filtering probability of the last changepoint location $C_{t}$ can be obtained via the forward recursions
\begin{equation}
\mathrm{Pr}(C_{t+1} = s | \bm{y}_{1:(t+1)}) \propto 
\begin{cases} W_{t+1}^{(s)} \mathrm{Pr}(C_{t+1} = s | C_{t} = s) \mathrm{Pr}(C_{t} = s | \bm{y}_{1:t}) , \;\;\; \forall \, s \in \mathcal{S}_{t}  \\
W_{t+1}^{(t)} \sum_{r=1}^{t-1} \mathrm{Pr}(C_{t+1} = t | C_{t} = r) \mathrm{Pr}(C_{t} = r | \bm{y}_{1:t}) , \;\;\; s=t \end{cases} \; ,
\label{eqn:filtering}
\end{equation}
where $\mathcal{S}_{t}$ (with $\mathcal{S}_{t} = \{t-1\} \bigcup \mathcal{S}_{t-1}$) is the collection of all possible changepoints before time $t$, and the weight is calculated as
\begin{equation}
W_{t+1}^{(s)} = f(y_{t+1} | C_{t+1} = s, \bm{y}_{1:t}) = \frac{\sum_{m} L(s, t+1, m) p_{m}}{\sum_{m} L(s, t, m) p_{m}} \; .
\label{eqn:weight}
\end{equation}
Here $p_{m}$ is the prior probability of the polynomial regression model of order $m$, and $\sum_{m} L(s, t+1, m) p_{m}$ is the likelihood after marginalising over prior model probabilities.

After calculating the filtering probabilities, it becomes straightforward to obtain the changepoints and model indices as the maximum \textit{a posteriori} (MAP) estimate via the Viterbi algorithm. Define $\mathcal{H}_{s}$ to be the MAP choice of changepoints and model orders up to time $s$ (i.e., the historical optimal path before $s$). Define
\begin{equation*}
P_{t}(s, m) = \mathrm{Pr}(C_{t}=s, M_{t}=m, \mathcal{H}_{s}, \bm{y}_{1:t})
\end{equation*}
and 
\begin{equation*}
P_{t}^{\text{MAP}} = \mathrm{Pr}(\text{changepoint at } t, \mathcal{H}_{t}, \bm{y}_{1:t}) \; .
\end{equation*}
For all $t=1, \cdots, n$, $s=0, \cdots, t-1$, and $m = 1, \cdots, q$, the above two probabilities can be calculated recursively as
\begin{equation}
P_{t}(s, m) = \{ 1 - G(t-s-1) \} L(s,t,m) p_{m} P_{s}^{\text{MAP}} \; ,
\label{eqn:MAP1}
\end{equation}
and
\begin{equation}
P_{t}^{\text{MAP}} = \max_{s, m} \left \{ \frac{P_{t}(s, m) g(t-s)}{1 - G(t-s-1)} \right \} \; .
\label{eqn:MAP2}
\end{equation} 
Alternatively, the changepoints can be simulated using the forward filtering probabilities and the transition probabilities, starting from the last time point and moving backwards. Unlike a typical hidden Markov model, the smoothing probabilities $\mathrm{Pr}(C_{t} | \bm{y}_{1:n})$ are not required in the backwards simulation of changepoints, because the segments before and after a changepoint are assumed to be independent. 

Note that the BOCPD method does not produce a full posterior of the changepoint locations and model types. The filtering distribution at time $t$ is only the posterior of the last possible changepoint before time $t$. However, by simulating a large number of configurations of changepoints from the filtering distributions and visualising the result as the count of each simulated changepoint out of the total number of configurations, one can get some intuition on the posterior distribution of changepoint locations.

\subsection{Challenges in parameter estimation} 

In real-world problems, the parameters of the observation models are typically unknown. Hence, being able to calculate the marginalised likelihood as in \eqref{eqn:intLike} becomes essential. This can be challenging when the parameter does not have a conjugate prior and when there is no analytical solution. Numerical integration is often required in this case. However, from empirical experiments, it appeared that the numerical integration for marginal likelihood does not always converge. As the unknown parameters are unique to each segment, it is also impossible to get any good estimation prior to the analysis, and using a global estimation often result in poor segmentation. This calls for method that can be implemented sequentially as the BOCPD algorithm moves forward.

The parameter estimation problem is not new in changepoint detection. However, existing literature focuses more on the estimation of the static parameters, such as prior parameters and model parameters that do not change with segments (e.g., if we assume $\theta_{j}$ in $\bm{\eta}_{j}$ from model (\ref{eqn:cptProblem}) to be fixed for $j = 1, \cdots, k$). For example, \cite{particleMCMC} proposed the particle Metropolis-Hastings and the particle Gibbs sampling methods to estimate the static parameters of the changepoint detection model; \cite{UncertaintyCpt} estimated the global parameters using sequential Monte Carlo; \cite{onlineEMCPT} proposed the use of an online EM algorithm \citep{onlineEM} to estimate the static parameters. \cite{DynamicCPT} carried out the inference of the segment-specific parameters alongside the inference of the run-length as an augmented state using particle filters. However, literature aiming at the segment-specific parameters is relatively limited. This might be explained by the fact that segment-specific parameters can sometimes be estimated post-segmentation, provided the evaluation of the likelihood can proceed without knowing their values. Overall, the above mentioned methods do not generalise to the parameter estimation problem (or the likelihood evaluation problem) in this study straightforwardly, some due to the difficulty in extending the methods to segment-specific parameters, some due to the challenge in incorporating multiple candidate models. 

In the following section, we propose two extensions of the BOCPD method to tackle the parameter estimation challenge from two different perspectives. The goal is to find ways to either approximate the marginal likelihood or obtain an estimation of the segment-specific parameter, so that the BOCPD method can be proceeded. The key idea is to combine the classic algorithm with additional steps that can process the unknown parameter sequentially during the forward filtering recursions. The first approach utilises particle filters to obtain a posterior distribution of the unknown parameters, which not only produces an estimation of the parameter, but also an approximation to the marginal likelihood. This extension is motivated by the sequential Monte Carlo method for estimating the time-varying parameters in \cite{TimeVarySMC}. The key to this extension is to treat the unknown parameter as part of the hidden state to be inferred during the recursion. The second approach aims to estimate the unknown parameters directly during the forward filtering recursion. The key to this extension is to combine the online gradient method with the BOCPD algorithm. The two approaches are introduced with full detail in sections \ref{sec:BOCPD+PF} and \ref{sec:BOCPD+OG}.

\section{Estimating segment specific parameters} \label{sec:Method}

Without loss of generality, assume model type $m^{\ast}$ has unknown parameters $\bm{\eta} = (\theta, \beta)$, where $\beta$ has a conjugate prior, denoted $\pi(\beta)$, but the prior distribution of $\theta$, denoted $\pi(\theta)$, does not have a conjugate prior. Assume that $\beta$ can be marginalised in closed form given $\theta$. In other words, although there is no analytical form for the marginal likelihood of model type $m^{\ast}$ as in formula (\ref{eqn:intLike}), a partially marginalised likelihood over $\beta$ can be obtained in closed form once $\theta$ is known. To emphasise the fact that $\theta$ changes from segment to segment, we use $\theta_{s}$ to refer to the parameter associated with the segment after the changepoint $s$ (i.e., $C_{t} = s$), and use $\theta_{st}$ to refer to the estimate of $\theta_{s}$ based the information up to time $t$ (i.e., based on $\bm{y}_{(s+1):t}$). In a typical recursion at time $t$, all $\theta_{s}$, with $s \in \mathcal{S}_{t}$, need to be evaluated. Finally, we will refer to the extension using particle filters as the BOCPD+PF method and the extension using online gradient descent as the BOCPD+OG method from here onward.

\subsection{Estimate the unknown parameter using a particle filter} \label{sec:BOCPD+PF}

The BOCPD+PF extension attempts to approximate the posterior distribution (or filtering distribution) of the unknown parameter $p(\theta_{s} | \bm{y}_{(s+1):t}, C_{t}=s, M_{t}=m^{\ast})$ using particle filters, and then an approximation to the marginal likelihood can be obtained. However, $\theta_{s}$ is a fixed parameter for the segment $\bm{y}_{(s+1):t}$. In order to apply the particle filter, one way is to treat the parameter as part of the hidden state and introduce some form of ``artificial evolution'' to the parameter. This idea was originally used in the inference of dynamic states, where small random disturbances are added to the state particles to reduce sample degeneracy, but it has since been used to estimate the fixed parameters in a dynamic model \citep{LWfilter, TimeVarySMC}, and hence the term ``artificial evolution''. 

For the parameter estimation problem in this case, a simple artificial evolution can be $\theta_{s(t+1)} = \mu(\theta_{st}) + \zeta_{t+1}$, where $\mu(\theta_{st})$ is some function of the previous state and $\zeta_{t} \sim \mathcal{N}(0, V_{t}) $ is some Gaussian distributed noise. In a typical recursion from time $t$ to $t+1$, each parameter associated with a potential last changepoint before $t+1$ will be updated according to the artificial evolution, with
\begin{equation}
\theta_{s(t+1)} = 
\begin{cases}
    \mu(\theta_{st}) + \zeta_{t+1}, \;\;\; \forall s \in \mathcal{S}_{t}\\
    \theta^{\ast} \sim \pi (\theta) , \;\;\; \text{for } s=t
\end{cases} .
\label{eqn:ThetaTransition}
\end{equation}
Here the first line corresponds to the case where $t$ is not a changepoint, i.e., $C_{t+1} = C_{t} = s$, $s \in \mathcal{S}_{t}$; the second line corresponds to the case where $t$ is a new changepoint, i.e., $C_{t+1} = t$, and hence we initialise the parameter using a random draw $\theta^{\ast}$ from the prior distribution $\pi(\theta)$. The evolution in (\ref{eqn:ThetaTransition}) propagates the particles, which are then used to approximate the posterior distribution and the marginal likelihood. Note that one may define an augmented state which includes both the last changepoint $C_{t}$, model indices $M_{t}$ and the parameters $\theta_{st}$, $s \in \mathcal{S}_{t}$. Although the unique recursion of BOCPD means that the augmented state does not enjoy a simple form as in \cite{TimeVarySMC} or \cite{DynamicCPT}. In fact, the content of the augmented state will change according to the collection of last changepoints $\mathcal{S}_{t}$. Therefore, we do not write explicitly the augmented state going forward.

What remains is to identify an appropriate form for the evolution in (\ref{eqn:ThetaTransition}). The simplest mean function would be $\mu(\theta_{st}) = \theta_{st}$, but this can lead to over-dispersion. Here we choose to use the Liu \& West filter \citep{LWfilter}, which is a special type of particle filter that approximates the target distribution via a mixture of Gaussian distributions. The mean and variance function of the propagation distribution is chosen to avoid information loss and over-dispersion due to the artificial evolution. Here we elaborate on some details. Define the samples, or particles, of the parameter $\theta_{s}$ at time $t$ as $\theta_{st}^{(i)}$, $i =1, \cdots, N_{\theta}$, and the mean and variance functions of the propagation distribution as $\mu(\cdot)$ and $V(\cdot)$. The Liu \& West filter approximates the posterior distribution of the difficult unknown parameter, as
\begin{equation*}
    p(\theta_{s} | \bm{y}_{(s+1):(t+1)}) \approx \sum_{i=1}^{N_{\theta}} w_{(t+1)}^{(i)} \mathcal{N} ( \mu(\theta_{st}^{(i)}) ,  V(\theta_{st}) )  \; ,
\end{equation*} 
where the mean and variance functions are
\begin{align*}
\mu(\theta_{st}^{(i)}) & = a \theta_{st}^{(i)} + (1-a) \bar{\theta}_{s} \\
V(\theta_{st}) & = \sum_{i=1}^{N_{\theta}} w_{t}^{(i)} (\theta_{st}^{(i)} - \bar{\theta})_{s} (\theta_{st}^{(i)} - \bar{\theta}_{s})^{\top} \; ,
\label{eqn:LWFmean}
\end{align*}
with a weighted mean, $\bar{\theta}_{s} = \sum_{i=1}^{N_{\theta}} w_{t}^{(i)}\theta_{st}^{(i)} $, shrinkage parameter $a$, and particle weights
\begin{equation*}
w_{t}^{(i)} \propto \frac{f(\bm{y}_{(s+1):t} | C_{t} = s, M_{t}=m^{\ast}, \theta_{st}^{(i)})}{f(\bm{y}_{(s+1):t} | C_{t} = s, M_{t}=m^{\ast}, \mu(\theta_{s(t-1)}^{(i)}) )} \; .
\label{eqn:LWFweight}
\end{equation*}

From time $t$ to $t+1$, we first sample the new generation of particles $\theta_{s(t+1)}^{(i)}$ from the propagation distribution $\mathcal{N} (\mu(\theta_{st}^{(i)}), h^{2} V(\theta_{st})) $, with shrinkage $h$ ($h = \sqrt{1-a^{2}}$). Next, we update the weights $w_{t+1}^{(i)}$ and re-sample the particles $\{\theta_{s(t+1)}^{(i)}\}_{i=1}^{N_{\theta}}$, where the probability of sampling $\theta_{s(t+1)}^{(k)}$ is proportional to $w_{t+1}^{(k)}$. Then we calculate the marginalised likelihood (over $\beta_{s})$ of model $m^{\ast}$ for the segment $\bm{y}_{(s+1):(t+1)}$ conditional on $\theta_{s(t+1)}^{(k)}$ as 
\begin{equation}
L(s, t+1, m^{\ast})^{(k)} = \int f(\bm{y}_{(s+1):(t+1)} | C_{t+1}=s, M_{t+1}=m^{\ast}, \theta_{s(t+1)}^{(k)}, \beta_{s}) \pi(\beta_{s}) \mathrm{d} \beta_{s}  \; .
\label{eqn:ptcLike}
\end{equation}
Finally, the likelihood after marginalising all unknown parameters is approximated using the resampled particles 
\begin{align}
L(s, t+1, m^{\ast}) & = f(\bm{y}_{(s+1):(t+1)} | C_{t+1}=s, M_{t+1}=m^{\ast}) \label{eqn:marginLike} \\
& \approx \frac{1}{N_{\theta}}\sum_{k=1}^{N_{\theta}} f(\bm{y}_{(s+1):(t+1)} | C_{t+1} = s, M_{t+1}=m^{\ast}, \theta_{s(t+1)}^{(k)}) = \frac{1}{N_{\theta}} \sum_{k=1}^{N_{\theta}} L(s, t+1, m^{\ast})^{(k)} \; . \nonumber
\end{align}
Once the marginal likelihood $L(s, t+1, m)$ for all models $m$ and last changepoints $s$ are evaluated, the weights $W_{t+1}^{(s)}$ as defined in \eqref{eqn:weight} can be obtained, and hence the filtering distribution $\mathrm{Pr}(C_{t+1} = s | \bm{y}_{1:(t+1)})$ can be calculated using \eqref{eqn:filtering}.

Before presenting the full procedure in Algorithm (\ref{alg:BOCPD_PF}), we define one more quantity that is useful in practice, which is the minimum segment length $d$. The BOCPD algorithm was introduced assuming $d=1$, i.e., the most recent possible changepoint before time $t$ is $t-1$. In practice, the minimum segment length can be any positive value that is reasonable to the problem. Sometimes, it is used to ensure that the model can be estimated (e.g., at least two observations are needed to estimate a mean and a variance); sometimes it is used to reflect the physical restriction of the problem. As a result, the most recent possible changepoint before $t$ given minimum segment length $d$ would be $t-d$.

\begin{algorithm}[htb]
\caption{The extended forward filtering recursion with particle filter}\label{alg:BOCPD_PF}
\begin{algorithmic}
\Require observation $\bm{y}_{1:n}$; model probability $p(m)$; run length distribution $g(t-s)$;  prior density $\pi(\theta)$; minimum segment length $d$
\Ensure sample $\theta^{\ast(i)} \sim \pi(\theta)$, $i = 1, \cdots, N_{\theta}$, and set $\theta_{0d}^{(i)} = \theta^{\ast(i)}$ 
\For{$t = d+1, \cdots, 2d-1$} 
  \State 1. update $\theta_{0(t-1)}^{(i)}$ through particle filter to obtain new particles $\theta_{0t}^{(k)}$, $k = 1, \cdots, N_{\theta}$
  \State 2. calculate likelihood $L(0, t, m^{\ast})$ using formulae (\ref{eqn:ptcLike}) and (\ref{eqn:marginLike}) with particles $\theta_{0t}^{(k)}$
  \State 3. calculate likelihood $L(0, t, m)$ for $m \neq m^{\ast}$ using formula (\ref{eqn:intLike})
  \State 4. set $\mathrm{Pr}(C_{t} = 0 | \bm{y}_{1:t}) = 1$ and $\mathcal{S}_{t} = \{ 0 \}$ \EndFor
\For{$t = 2d, \cdots, n$} 
  \For{$s \in \mathcal{S}_{t-1} $}  \Comment{possible last changepoints kept from previous recursion}
    \State (1). update $\theta_{s(t-1)}^{(i)}$ through particle filter to obtain new particles $\theta_{st}^{(k)}$, $k = 1, \cdots, N_{\theta}$
    \State (2). calculate likelihood $L(s, t, m^{\ast})$ using formulae (\ref{eqn:ptcLike}) and (\ref{eqn:marginLike}) with particles $\theta_{st}^{(k)}$ 
    \State (3). calculate likelihood $L(s, t, m)$ for $m \neq m^{\ast}$ using formula (\ref{eqn:intLike}) \EndFor
  \For{$s = t-d $}  \Comment{newly added last changepoint at $t-d$}  
    \State (1). sample $\theta^{\ast(i)} \sim \pi(\theta)$, $i = 1, \cdots, N_{\theta}$, and set $\theta_{st}^{(i)} = \theta^{\ast(i)}$ 
    \State (2). calculate likelihood $L(s, t, m^{\ast})$ using formulae (\ref{eqn:ptcLike}) and (\ref{eqn:marginLike}) with particles $\theta_{st}^{(k)}$
    \State (3). calculate likelihood $L(s, t, m)$ for $m \neq m^{\ast}$ using formula (\ref{eqn:intLike})  \EndFor
  \State 1. calculate the weights $W_{t}^{(s)}$, $s \in \mathcal{S}_{t-1} \bigcup \{ t-d \}$, using formula (\ref{eqn:weight}) 
  \State 2. calculate the filtering distribution $\mathrm{Pr}(C_{t} = s | \bm{y}_{1:t})$, $s \in \mathcal{S}_{t-1} \bigcup \{ t-d \}$, using formula (\ref{eqn:filtering})
  \State 3. set $\mathcal{S}_{t} = \mathcal{S}_{t-1} \bigcup \{ t-d \}$
\EndFor \\
\Return filtering distribution $\mathrm{Pr}(C_{t} | \bm{y}_{1:t}) $ for all $t=d+1, \cdots, n$; posterior density of the unknown parameter $\theta_{s}$ for all possible $s$. 
\end{algorithmic}
\end{algorithm}

Compared to the original BOCPD method in \cite{BOCPD}, this extension will increase the computational cost, as the number of particles required in the particle filter is usually large. This will be discussed in detail in Section \ref{sec:ComputingCost}. However, the particle filter have some clear advantages over the numerical integration version of the BOCPD as it provides a posterior distribution of the unknown parameter and it largely avoids the convergence problem of the numerical integration.

\subsection{Estimate the unknown parameter using online gradient descent} \label{sec:BOCPD+OG}

This extension tackles the difficult unknown parameter problem from a different perspective. Instead of obtaining an approximation of the posterior distribution of the parameter to evaluate the marginal likelihood, it choose to estimate the unknown parameter directly on the go. Although the result is just a point estimate without uncertainty measure, it is computationally more efficient, and hence would be suitable for some applications that involve very long time series

Gradient descent is one of the most commonly used methods to solve an optimisation problem. Two types of gradient descent methods are available: batch gradient descent, which uses the entire data set at each update step, and stochastic gradient descent, which drastically simplifies the batch method by randomly selecting a data point $y_t$ at each step to perform the update \citep{StochasticGradient, GradientOverview}. Online gradient (OG) descent is a special case of stochastic gradient descent. Instead of randomly selecting an observation, the observation is picked sequentially over time. A generic update step in online gradient descent for parameter $\theta_{s}$ can be written as 
\begin{equation*}
    \theta_{s(t+1)} = \theta_{st} - \gamma_{t} \nabla_{\theta} \, Q(y_{t}, \theta_{st})
\end{equation*}
for some target function $Q(y_{t}, \theta_{st})$ and step size (or learning rate) $\gamma_{t}$, which may or may not change with time $t$. Compared to batch gradient descent, the convergence rate of online gradient descent is limited by the noisy approximation of the true gradient. However, by appropriately choosing the step size, similar convergence rates can be achieved \citep{StochasticGradient, SGDtricks}. 

Similar to the particle filtering update in (\ref{eqn:ThetaTransition}), when a time point $s$ is first recognised as a potential last changepoint prior to $t$, we initialise $\theta_{s}$ as a random draw $\theta^{\ast}$ from the the prior distribution $\pi(\theta)$, and set $\theta_{st} = \theta^{\ast}$. Then, at each recursion, the parameter estimation is updated as follows
\begin{equation}
\theta_{s(t+1)} = \theta_{st} - \gamma_{t}  \nabla_{\theta_{s}} \, l(\theta_{st}; \, y_{t}) 
\label{eqn:gradient1}
\end{equation}
where $l(\theta_{s}; y_{t}) = \log f(y_{t} | C_{t}=s, M_{t}=m^{\ast}, \theta_{s})$ is the log-likelihood function of $\theta_{s}$ after integrating out $\beta_{s}$ and $\nabla_{\theta_{s}} \, l(\theta_{st}; \, y_{t}) $ is its gradient function, or if a second order gradient scheme is used, 
\begin{equation}
\theta_{s(t+1)} = \theta_{st} - \gamma_{t} H_{st}^{-1} \nabla_{\theta_{s}} \, l(\theta_{st}; \, y_{t} ) \; ,
\label{eqn:gradient2}
\end{equation}
where $H_{st}^{-1}$ is a positive definite matrix that approximates the inverse of the Hessian of the log-likelihood function. Unlike in batch gradient descent, where the second order update shows a faster convergence rate than the first order update, the second order update does not necessarily result in better convergence in online gradient descent \citep{SGDtricks}. Therefore, the choice between the two update schemes may need to be investigated case by case in practice, along with the step size. 

Returning to the original BOCPD problem, during the forward filtering recursion from time $t$ to $t+1$, we update the estimation of $\theta_{s}$, for $s \in \mathcal{S}_{t}$, from $\theta_{st}$ to $\theta_{s(t+1)}$ using either the first order gradient update (\ref{eqn:gradient1}) or the second order update (\ref{eqn:gradient2}), and we initialise the unknown parameter associated with the newly added last changepoint $s=t$ during the current recursion. The updated parameters are then used to evaluate the marginal likelihood $L(s, t+1, m^{\ast})$ and to calculate the weights of the potential last changepoints, $W_{t+1}^{(s)}$. Again the parameters that have conjugate priors are marginalised conditioning on $\theta_{s(t+1)}$, giving
\begin{equation}
L(s, t+1, m^{\ast}) = \int f(\bm{y}_{(s+1):(t+1)} | C_{t+1}=s, M_{t+1}=m^{\ast}, \theta_{s(t+1)}, \beta_{s}) \pi(\beta_{s}) \mathrm{d} \beta_{s}  \; .
\label{eqn:grdLike}
\end{equation}
The extended forward filtering recursion algorithm with gradient descent is given in Algorithm (\ref{alg:BOCPD_OG}).

\begin{algorithm}[htb]
\caption{The extended forward filtering recursion with online gradient descent}\label{alg:BOCPD_OG}
\begin{algorithmic}
\Require observation $\bm{y}_{1:n}$; model probability $p(m)$; run length distribution $g(t-s)$;  prior density $\pi(\theta)$; minimum segment length $d$
\Ensure sample $\theta^{\ast} \sim \pi(\theta)$ and set $\theta_{0d} = \theta^{\ast}$
\For{$t = d+1, \cdots, 2d-1$} 
  \State 1. calculate the gradient $\nabla_{\theta_{0}} \, l(\theta_{0(t-1)}; \, y_{d})$ or the inverse Hessian $H_{0(t-1)}^{-1}$, update $\theta_{0t}$ from $\theta_{0(t-1)}$ through gradient descent (\ref{eqn:gradient1}) or (\ref{eqn:gradient2}) 
  \State 2. calculate likelihood $L(0, t, m^{\ast})$ using formula (\ref{eqn:grdLike}) with $\theta_{0t}$
  \State 3. calculate likelihood $L(0, t, m)$ for $m \neq m^{\ast}$ using formula (\ref{eqn:intLike})
  \State 4. set $\mathrm{Pr}(C_{t} = 0 | \bm{y}_{1:t}) = 1$ and $\mathcal{S}_{t} = \{ 0 \}$ 
\EndFor
\For{$t = 2d, \cdots, n$} 
  \For{$s \in \mathcal{S}_{t-1} $}  \Comment{possible last changepoints kept from previous recursion}
    \State (1). calculate the gradient $\nabla_{\theta_{s}} \, l(\theta_{s(t-1)}; \, y_{d})$ or the inverse Hessian $H_{s(t-1)}^{-1}$, update $\theta_{st}$ from $\theta_{s(t-1)}$ through gradient descent (\ref{eqn:gradient1}) or (\ref{eqn:gradient2}) 
    \State (2). calculate likelihood $L(s, t, m^{\ast})$ using formula (\ref{eqn:grdLike}) with $\theta_{st}$
    \State (3). calculate likelihood $L(s, t, m)$ for $m \neq m^{\ast}$ using formula (\ref{eqn:intLike}) \EndFor
  \For{$s = t-d $}  \Comment{newly added last changepoint at $t-d$}  
    \State (1). sample $\theta^{\ast} \sim \pi(\theta)$ and set $\theta_{st} = \theta^{\ast}$
    \State (2). calculate likelihood $L(s, t, m^{\ast})$ using formula (\ref{eqn:grdLike}) with $\theta_{st}$
    \State (3). calculate likelihood $L(s, t, m)$ for $m \neq m^{\ast}$ using formula (\ref{eqn:intLike})  \EndFor
  \State 1. calculate the weights $W_{t}^{(s)}$, $s \in \mathcal{S}_{t-1} \bigcup \{ t-d \}$, using formula (\ref{eqn:weight}) 
  \State 2. calculate the filtering distribution $\mathrm{Pr}(C_{t} = s | \bm{y}_{1:t})$, $s \in \mathcal{S}_{t-1} \bigcup \{ t-d \}$, using formula (\ref{eqn:filtering})
  \State 3. set $\mathcal{S}_{t} = \mathcal{S}_{t-1} \bigcup \{ t-d \}$
\EndFor \\
\Return filtering distribution $\mathrm{Pr}(C_{t} | \bm{y}_{1:t}) $ for all $t=d+1, \cdots, n$; posterior density of the unknown parameter $\theta_{s}$ for all possible $s$. 
\end{algorithmic}
\end{algorithm}

One of the crucial choices in the online gradient descent method is the step size $\gamma_{t}$, which can affect the convergence of the method \citep{SGDtricks, GradientOverview}. Extensive research has been carried out to investigate the properties of different step size schedules and to find the appropriate learning rate for different scenarios (e.g., \cite{LearningRateAdapt, AdaptiveGrd, DOG, coullon2023efficient}), as well as recent work on learning-rate-free approaches (e.g., \cite{orabona2016coin,orabona2020tutorial,sharrock2023coin,sharrock2023learning}). Choices range from a fixed step size, which is not optimal but easy to implement, and a diminishing step size (such as $\gamma_{t} \sim 1/t$), which is optimal under certain conditions \citep{SGDtricks}, to an adaptive step size, which usually results in better convergence, but requires additional calculation per iteration. Here, an adaptive step size schedule will be used, and to avoid additional choices in an already complicated application problem, we choose to use a parameter-free approach, the ``distance over gradient'' (DOG) method by \cite{DOG}. Thus the step size $\gamma_{t}$ is 
\begin{equation}
    \gamma_{t} = \frac{\max_{j(j<t)} || \theta_{sj} - \theta_{st} ||}{\sqrt{\sum_{j<t} || \nabla_{\theta_{s}} \, l(\theta_{sj}; \, y_{j} , \cdots ) ||^{2} } } \, ,
\end{equation}
with the initial step $\gamma_{0} = r_{\epsilon} / \nabla_{\theta_{s}} \, l(\theta_{s(s+1)}; \, y_{s+1} , \cdots ) $ for some small $r_{\epsilon}$. The optimality gap bound of DOG is shown to be double-logarithmic in $t$ and logarithmic in $1/r_{\epsilon}$ \citep{DOG,RDOG}.

Although the two extensions in this section were introduced here using a particular type of particle filter or a particular gradient update formula, the method itself may be more flexible, as the BOCPD algorithm, which sits at its core, is a generic method. For both of the extensions, different configurations of the algorithm may be considered. For example, one can choose which parameters to update during the filtering recursion and tailor the algorithm to suit an update of multiple parameters. For the BOCPD+OG extension, this could mean using different step sizes for different parameters in order to achieve better convergence. The choice of particle filter is also not restricted to the Liu \& West filter. A wide range of choices is available in the literature, from generic filters such as sequential importance resampling and the auxiliary particle filter to more specific ones developed for particular problems \citep{OnSMC, TutorialPF, ReviewPF}.

\section{Practical considerations} \label{sec:Practical}

Before applying the extended algorithms to the real soil moisture time series data, some implementation details of the method are discussed here. These include the ways of reducing computational cost, and the selection of priors, tuning parameters. The proposed extensions are then illustrated on simulated time series data. The main aim of the illustration is to show the overall performance as well as the different features of the two extensions.

\subsection{Reducing computational cost} \label{sec:ComputingCost}

Computational cost is a crucial consideration in real-world application problems. For example, in the modelling of soil moisture time series, the main concern is the increasing length of the time series $n$. The original BOCPD algorithm has a computational cost that is quadratic in $n$. This is because at each time step $t$, given a minimum segment length $d$, there are $t-d$ potential last changepoints to process. Therefore, running the forward filtering recursion from time 1 to $n$ requires $\mathcal{O} (n^{2})$ computational time. For the BOCPD+OG extension, the computational cost remains $\mathcal{O} (n^{2})$, but for the BOCPD+PF extension, this increases to $\mathcal{O} (n^{2} N_{\theta} )$ due to the use of $N_{\theta}$ particles. 

To reduce the computational cost, \cite{BOCPD} applies an additional resampling step to keep a smaller number of potential last changepoints at each iteration. In particular, a stratified optimal re-sampling (SOR) method is introduced in \cite{BOCPD} and \cite{PFresample}, which aims to reduce the overall computational cost to $\mathcal{O} (n P N_{\theta} )$, where $P$ is the maximum number of potential last changepoints to keep after re-sampling at each time step. The resampling involves the following steps. For the maximum number of changepoints $P$, (i) calculate a threshold $\alpha$, such that $\sum_{i} \min \{ 1, W_{t}^{(s)}/\alpha \} = P$, (ii) keep all changepoints $s$ such that $W_{t}^{(s)} \geq \alpha$, (iii) assuming there are $A$ changepoints kept from step (ii), use stratified re-sampling to sample $P-A$ changepoints from the remaining changepoints, and assign each with weight $\alpha$. Details of the re-sampling are given in \cite{BOCPD} and \cite{PFresample}. 

The implementation of the SOR requires a decision on the maximum number of potential last changepoints $P$. To some extent, this is a trade-off between accuracy and computational cost, i.e., the larger the $P$ is, the more accurate the result would be, but the higher the computational cost. Whilst this is less influential for BOCPD+OG, it can make a big difference to the computational cost of BOCPD+PF as it usually requires a large number of particles $N_{\theta}$ for each changepoint $s$.

\subsection{Selecting priors and tuning parameters} \label{sec:Configuration}

The extended BOCPD algorithm relies on a few global parameters to run, for example, the parameters of the run-length distribution and prior distributions, and tuning parameters such as the minimum segment length. They need to be determined in advance, and empirical experiments showed that they can have some impact on the result. Therefore, a careful selection of these parameters could be beneficial to particular application problem. 

The run-length distribution determines the transition probability, which affects the posterior distributions of the changepoints. As seen from the MAP estimation of changepoints (recall \eqref{eqn:MAP1} and \eqref{eqn:MAP2}), the run-length distribution can have a direct impact on result. For example, although a geometric distribution gives constant transition probability, the probability of having a changepoint increases with the run-length, resulting in an increasingly large impact on the posterior distribution. Some selection procedure using a grid search may be helpful to obtain a suitable run-length distribution. Alternatively, a particle Metropolis-Hastings (PMH) algorithm as introduced in \cite{particleMCMC} may be applied to identify an appropriate run-length distribution parameter. The PMH algorithm can be combined with both the extensions, although the additional computational cost it introduces would make its integration with the BOCPD+PF extension very challenging. 

The prior distributions of the models and parameters can also have an impact. In practice, the prior model probability may be determined based on background knowledge about the time series (e.g., if one type of mode is likely to occur more frequently than another) or the preference for simpler models over more complicated ones. In the absence of such information, it makes sense to assign equal probability to each candidate model. The prior probabilities on the parameters may be determined based on background knowledge. For example, the unknown parameter may have a clear physical meaning, which suggests the use of certain types of distribution. Otherwise, generic non-informative priors can be used. The prior distribution of the residual variance $\sigma^{2}$ can have some impact on changepoint detection. In some initial experiments, when the inverse Gamma prior was assigned to $\sigma^{2}$, various choices of parameters in the inverse Gamma distribution were investigated. It appeared that the number of changepoints detected is quite sensitive to the choice of parameters. When the high-density region of the inverse Gamma distribution is concentrated around smaller $\sigma^{2}$, the algorithm tends to detect more changepoints, and vice versa. Therefore, some sensitivity analysis may be needed when applying the method. In practice, a useful starting point is to use a prior distribution that reflects the scale of the model residuals, which can be estimated from a pre-run of the algorithm on some sample data. 

As mentioned at the end of section \ref{sec:BOCPD+PF}, a minimum segment length may be specified, whether it is for model identification purpose or for application-based reasons. For the soil moisture drydown analysis, a minimum length of a few days is often used to identify the drydown curves in literature. Hence we will follow this convention. Finally, with the re-sampling approach to reduce computational cost, there is a risk of a potential last changepoint being removed too early, because the algorithm for estimating the model parameters has not reached a result close enough to the true value. One way to reduce this risk is to keep the most recent changepoints for a period of time before they are allowed to be removed during re-sampling. This might be especially helpful for the extension using online gradient descent.

\subsection{Illustration of the method on simulated data} \label{sec:Illustration}

In this section, the proposed extensions were illustrated on various simulated time series examples. In particular, four scenarios were considered. They are,

\begin{enumerate}
\item[\textbf{S1:}] constant mean and exponential decay. Two types of candidate models are considered, (1) the exponential decay model $y_{t} = \beta_{0} + \beta_{1} \exp [ - \exp (\theta) (t - \tau)] + \epsilon_{t}$ and (2) the constant mean model $y_{t} = \alpha_{0} + \epsilon_{t}$. The difficult parameter is the decay parameter $\theta$ in the exponential decay model. Three changepoints were located at $t = 205, 489, 782$, for a time series of length 1000, and the model indices of the four segments are (2, 1, 1, 1) respectively.

\item[\textbf{S2:}] linear trend and exponential decay. Two types of candidate models are considered, (1) the exponential decay model $y_{t} = \beta_{0} + \beta_{1} \exp [ - \exp (\theta) (t - \tau)] + \epsilon_{t}$ and (2) the linear trend model $y_{t} = \alpha_{0} + \alpha_{1} (t - \tau) + \epsilon_{t}$. The difficult parameter is the decay parameter $\theta$ in the exponential decay model. Three changepoints were located at $t = 252, 524, 766$, for a time series of length 1000, and the model indices of the four segments are (2, 1, 2, 1) respectively.

\item[\textbf{S3:}] constant mean and periodicity. Two types of candidate models are considered, (1) the periodic model $y_{t} = \beta_{0} + \beta_{1} \sin (\frac{x - \tau}{\theta}) + \epsilon_{t}$ and (2) the mean model $y_{t} = \alpha_{0} + \epsilon_{t}$.  The difficult parameter is the cycle parameter $\theta$ in the periodic model. Three changepoints were located at $t = 259, 534, 726$, for a time series of length 1000, and the model indices of the four segments are (2, 1, 2, 1) respectively.

\item[\textbf{S4:}] linear trend and periodicity. Two types of candidate models are considered, (1) the periodic model $y_{t} = \beta_{0} + \beta_{1} \sin (\frac{x - \tau}{\theta}) + \epsilon_{t}$, where $\tau$ is a changepoint, and (2) the linear trend model $y_{t} = \alpha_{0} + \alpha_{1} (t - \tau) + \epsilon_{t}$. The difficult parameter is the cycle parameter $\theta$ in the periodic model. Three changepoints were located at $t= 221, 528, 765$, for a time series of length 1000, and the model indices of the four segments are (2, 1, 2, 1) respectively. 

\end{enumerate}

\begin{figure}[!htb]
\begin{minipage}{0.5\textwidth}
\centering
\includegraphics[width=3.2in]{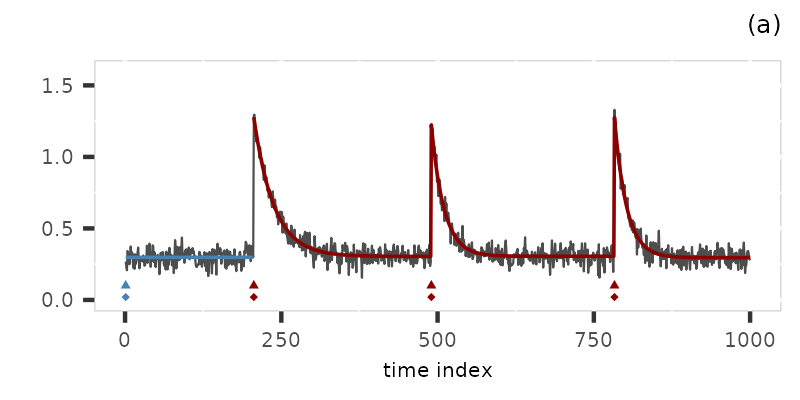}
\includegraphics[width=3.2in]{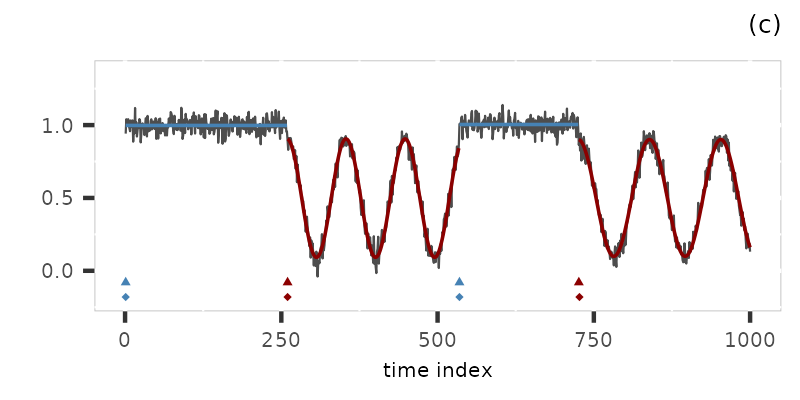}
\end{minipage} %
\begin{minipage}{0.5\textwidth}
\includegraphics[width=3.2in]{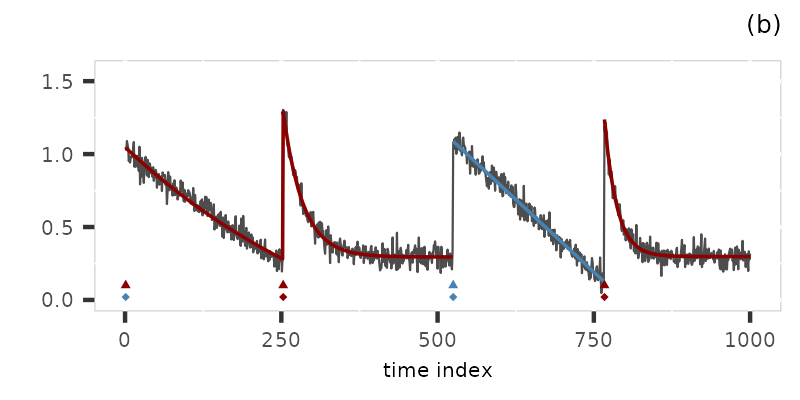}
\includegraphics[width=3.2in]{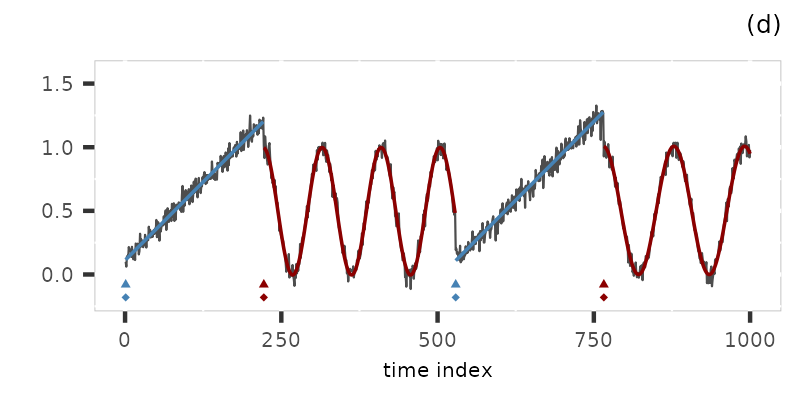}
\end{minipage} 
\captionsetup{labelfont=bf, font=small}
\caption{Simulated time series from scenarios (a) S1, (b) S2, (c) S3 and (4) S4, along with the identified changepoints (coloured triangles), the true changepoints (coloured rhombi), and the fitted segments (coloured curves) using parameters estimated from BOCPD+PF. }
\label{fig:ScenariosFitPF}
\end{figure}

\begin{figure}[!htb]
\begin{minipage}{0.5\textwidth}
\centering
\includegraphics[width=3.2in]{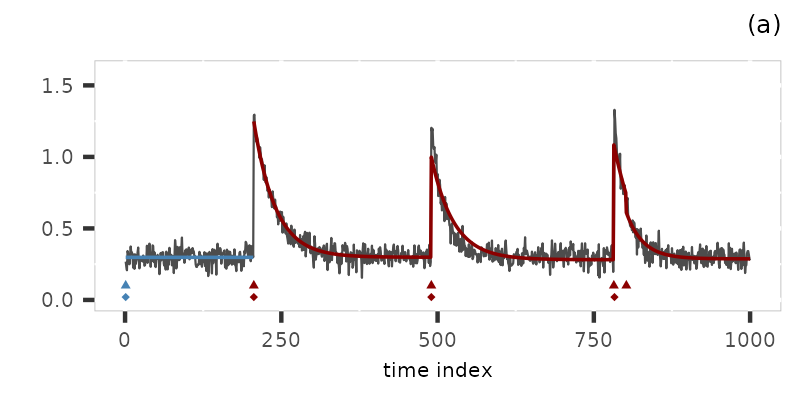}
\includegraphics[width=3.2in]{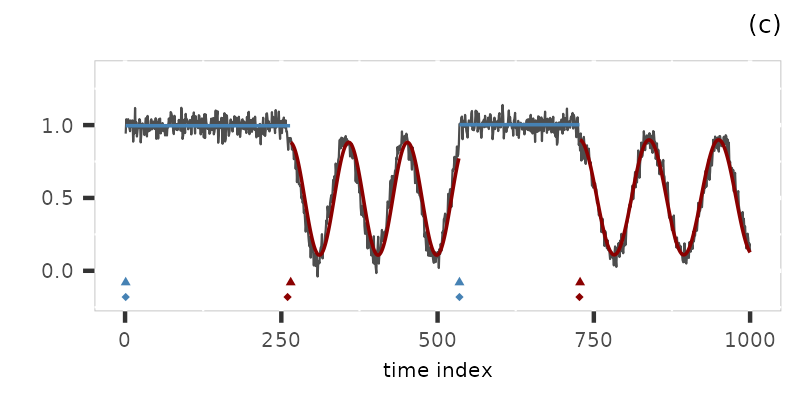}
\end{minipage} %
\begin{minipage}{0.5\textwidth}
\includegraphics[width=3.2in]{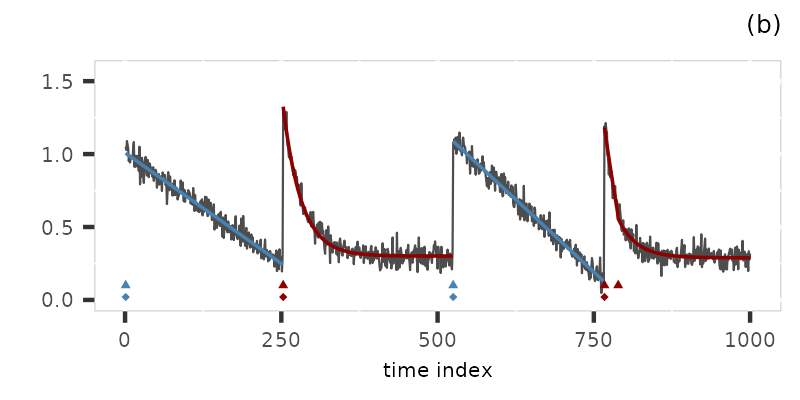}
\includegraphics[width=3.2in]{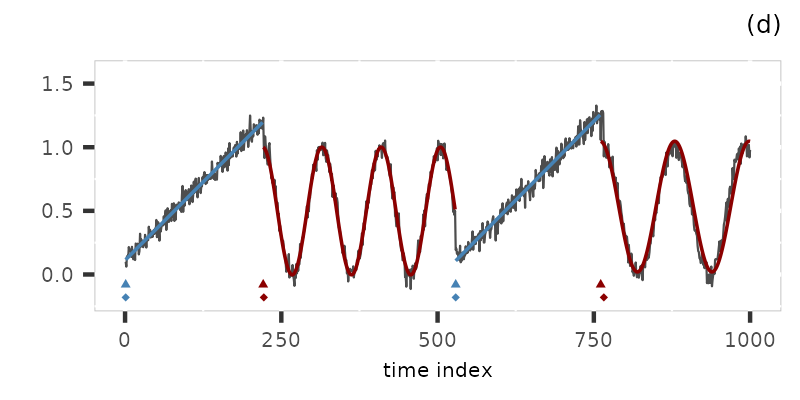}
\end{minipage} 
\captionsetup{labelfont=bf, font=small}
\caption{Simulated time series from scenarios (a) S1, (b) S2, (c) S3 and (4) S4, along with the identified changepoints (coloured triangles), the true changepoints (coloured rhombi), and the fitted segments (coloured curves) using parameters estimated from BOCPD+OG. }
\label{fig:ScenariosFitOG}
\end{figure}

The simulated time series are presented in Figures (\ref{fig:ScenariosFitPF}) and (\ref{fig:ScenariosFitOG}) as the black curves. For the illustration, both extensions introduced in Sections \ref{sec:BOCPD+PF} and \ref{sec:BOCPD+OG} are applied to the simulated time series. That is, (i) BOCPD+PF, which uses a Liu \& West filter for the difficult parameter with 1000 particles ($N_{\theta} = 1000$), and (ii) BOCPD+OG, which uses second order online gradient descent to estimate the difficult parameter, where the step size is determined via the ``distance over gradient'' (DOG) method \citep{DOG}. The original BOCPD with numerical integration for all parameters was also implemented, whenever the numerical integration converges with a maximum of 1000 subdivisions, for reference. Other implementation details include: (a) all models are assumed to be equally likely, (b) normal priors are given to the model coefficients $\alpha_{0}$, $\alpha_{1}$, $\beta_{0}$, $\beta_{1}$ and inverse Gamma priors are given to the model residuals $\epsilon_{t}$, which results in closed form solutions to the marginal likelihood given the difficult parameter, (c) run-length distribution parameter is fixed to 0.005, (d) whenever the number of potential last changepoints goes above 80, re-sampling is applied to keep only 40 potential last changepoints, (e) for the step size parameter in DoG, $r_{\epsilon}$, three values, $1 \times 10^{-6}$, $5 \times 10^{-6}$ and $1 \times 10^{-7}$, are investigated, and the best output in terms of changepoint detection is reported.

\begin{table}[htb]
\centering
\small
\begin{tabular}{ll|ccc}
  \hline
 & & BOCPD+PF & BOCPD+OG & Original \\ 
  \hline
\textbf{S1} &  True positive & 1.00 & 1.00 & 1.00 \\ 
  & Precision & 1.00 & 0.80 & 1.00 \\ 
  & Model selection & 1.00 & 1.00 & 1.00 \\ 
   \hline
\textbf{S2} &  True positive & 1.00 & 1.00 & 1.00 \\ 
  & Precision & 1.00 & 0.80 & 1.00 \\ 
  & Model selection & 0.75 & 1.00 & 1.00 \\
   \hline
\textbf{S3} &  True positive & 1.00 & 1.00 & NA  \\ 
  & Precision & 1.00 & 1.00 & NA  \\ 
  & Model selection & 1.00 & 0.99 & NA \\
   \hline
\textbf{S4} &  True positive & 1.00 & 1.00 & NA \\ 
  & Precision & 1.00 & 1.00 & NA  \\ 
  & Model selection & 1.00 & 0.99 & NA \\
   \hline
\end{tabular}
\captionsetup{labelfont=bf, font=small}
\caption{Summary of the changepoint detection and model selection results via maximum \textit{a posteriori} from four illustrative example. In accuracy of model selection is calculated as the number of time points that are assigned the correct model divided by the length of the time series.} 
\label{tab:illustration}
\end{table}

After running the forward filtering algorithms, we obtained the MAP estimation of changepoints and model indices via the Viterbi algorithm. The true positive rate (the number of corrected identified changepoints divided by the number of true changepoints), and precision (the number of corrected identified changepoints divided by the number of identified changepoints) are summarised in Table \ref{tab:illustration}. In particular, we consider any detected changepoints that are less than 10 time points away from a true changepoint as a correctly identified changepoint. To investigate the performance of the algorithms in model selection, a rate of correct assignment was calculated as the number of time points that are assigned with the correct model index over the length of the time series. The result is also shown in Table \ref{tab:illustration}. As mentioned in section \ref{sec:ReviewBOCPD}, backward simulation procedure can provide an impression of the posterior distributions of the changepoint locations. We provide an illustration of the backward simulation in the supplemental document. 

It can be seen from the table that BOCPD+PF achieved high accuracy in identifying the changepoints in all four scenarios; whereas the BOCPD+OG tends to estimate more changepoints, resulting in a lower precision. The original BOCPD with numerical integration show high accuracy in detecting changepoints in S1 and S2, but it failed to converge in S3 and S4 with 1000 subdivisions in the numerical integration, and hence the results are reported as ``NA''. Although BOCPD+OG estimated spurious changepoints, its performance in model selection is comparable to BOCPD+PF. In terms of parameter estimation, when the changepoints are identified correctly, the estimation of the difficult unknown parameters $\theta$ via both BOCPD+PF (as posterior mean) and BOCPD+OG (as point estimation) are close to the true value, with the parameters estimated from BOCPD+PF being more accurate. For example, the true cycle parameters in S4 are 15 and 18 for two cosine segments. The posterior means from BPCOD+PF are 15.014 and 18.003, and the estimates from BOCPD+OG are 15.100 and 18.997. 

Finally, we present the identified changepoints, model selection result, and fitted segments using the estimated $\theta$ from the two algorithms fin Figures \ref{fig:ScenariosFitPF} and \ref{fig:ScenariosFitOG} respectively. The colour code used in the graphs is red for the model containing $\theta$ and blue for the alternative model. The triangles represent the estimated changepoints and the rhumbi represent the true changepoints. Whilst the blue curves do not require the information of $\theta$, the red curves rely heavily on the estimated $\theta$. Hence a close fit of the red curves to the simulated time series would indicate a good estimation of $\theta$. Overall, BOCPD+PF appeared to have better performance over BOCPD+OG on the simulated problems. However, the estimated parameters from BOCPD+OG are still reasonable. Considering the substantially smaller computational cost of BOCPD+OG compared to BOCPD+PF, it can still be a useful tool when modelling long time series.

\section{Application to soil moisture data from HOAL and NEON} \label{sec:Application}

In this section, the extended BOCPD algorithms are applied to different soil moisture data to segment the time series and investigate the drydown characteristics. In the first application, the BOCPD+OG extension is applied to long-term soil moisture time series from field sites HOAL 01 and HOAL 05 from the HOAL data portal \citep{HOAL} to investigate the temporal patterns. Both time series have a frequency of every two hours, covering a period from 2013-09-01 00:00:00 to 2015-04-15 05:00:01 (length = 6957) for HOAL 01 and from 2013-09-01 00:00:00 to 2015-03-19 05:00:00 (length = 6638) for HOAL 05. Although the soil moisture data from HOAL are available from August 2013 to December 2019, there are some major gaps over the years. As a result, the changepoint analyses were only conducted on the period from the selected periods, before the first major missing gap came in spring 2015. 

In the second application, soil moisture time series from field sites ONAQ and CPER from the NEON data portal \citep{NEONsoil} are modelled to investigate the drydown parameters. To achieve better parameter estimation, the BOCPD+PF extension is considered. The method is applied to the down-sampled time series, with down-sampling carried out to reduce the computational cost. In particular, the soil moisture time series from CPER was down-sampled to every four hours, covering a period from 2021-04-04 20:00:00 to 2021-10-30 04:00:00 (length = 1251); the soil moisture time series from TALL was down-sampled to every two hours, covering a period from 2018-02-21 19:30:00 to 2018-06-05 23:30:00 (length = 1251). 



\subsection{HOAL soil moisture time series}

\begin{figure}[!htb]
\begin{center}
\includegraphics[width=5.6in]{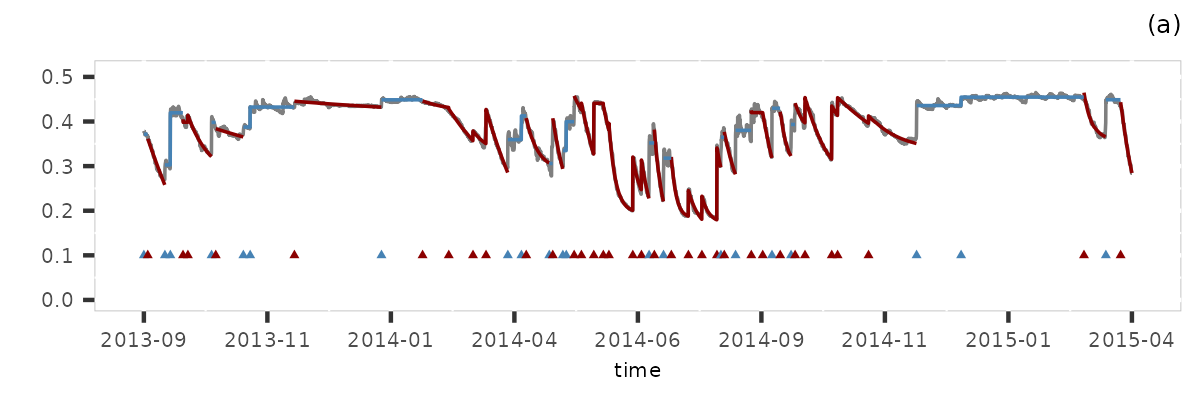}
\includegraphics[width=5.6in]{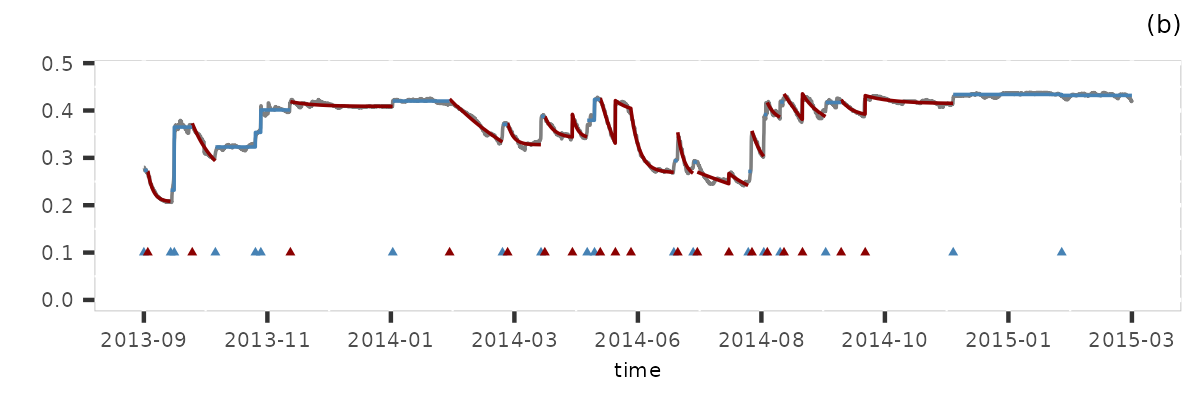}
\captionsetup{labelfont=bf, font=small}
\caption{The soil moisture time series from (a) HOAL 01 site and (b) HOAL 05 site, with the estimated changepoints and the fitted segments. The colour red represents the exponential decay model, and the colour blue represents the mean model.}
\label{fig:ApplicationHOAL}
\end{center}
\end{figure}

The Hydrological Open Air Laboratory (HOAL) in Petzenkirchen, Lower Austria, is a research catchment that has been established to advance the understanding of water-related flow and transport processes in the landscape \citep{HOAL}. The climate can be characterised as humid, with a mean annual temperature of $9.5^{\circ}\mathrm{C}$ and a mean annual precipitation of 823 mm. Precipitation tends to be higher during the summer than in the winter. The soil moisture time series recorded at some of these field sites have clear winter-summer shifts, for example, HOAL 01 and 05 sites (see Figure \ref{fig:ApplicationHOAL}). Whilst the summer time series shows typical drydown patterns, the winter time series shows more of a fluctuation around a high level of soil moisture content. This application aims to encode the changing patterns in the soil moisture time series and to investigate the rate of soil moisture drydown in the summer season. Two time series, from HOAL 01 site and HOAL 05 site, were investigated. Considering the lengths of the time series, we choose to use the BOCPD+OG extension in this application. 

Two candidate models, (a) an exponential decay model, $y_{t} = \beta_{0} + \beta_{1} \exp [ - \exp (\theta) (t - \tau)] + \epsilon_{t}$, and (b) a mean model, $y_{t} = \alpha_{0} + \epsilon_{t}$, are used to describe the patterns in the time series. The two models are assumed to have equal probability. The re-parameterisation $\exp [ - \exp (\theta) (t - \tau)]$ is used to avoid putting constraints on the parameter $\theta$. Obtaining the e-folding decay time scale, $\omega$, as in drydown model (\ref{eqn:Drydown}) from $\theta$ is straightforward. A geometric run-length distribution is assumed, with a probability of 0.005 that a time point is a changepoint. The two models were assumed to be equally likely, that is $p_{m}=0.5$. Initial values and prior distributions of the parameters were selected based on the characteristics of the data. In initial experiments, using shorter time series, different parameters in the run-length distribution, the inverse-Gamma prior distribution of $\sigma^{2}$, and the step size of the online gradient descent were investigated. It appears that the changepoint detection results are most sensitive for the parameters of the inverse-Gamma prior. An inverse Gamma distribution giving higher prior density on smaller $\sigma^{2}$ results in more changepoints. Therefore, we chose to use inverse Gamma priors that reflect the magnitude of the residuals based on empirical information. To avoid unrealistic estimation of the linear coefficients, the truncated Normal distribution was used as a prior distribution for the linear coefficients $\beta_{0}$ and $\beta_{1}$ in the exponential decay model. As this changes the typical Normal-inverse Gamma prior combination for linear regression models, some approximation were made to the integration. Some detail is given in the supplemental document. 

In this analysis, the BOCPD+OG algorithm was first run on four sets of parameters of the inverse-Gamma priors, all reflecting the magnitude of the empirical residuals, but in increasing order. Then the most appropriate model was chosen based on the BIC score and the closeness of the fit. The model that has fewer segments with an increasing trend is favoured. Note that although the truncated Normal priors can reduce the chance of getting an increasing segment, occasionally this could still happen. The results are shown in Figure \ref{fig:ApplicationHOAL}. 

Overall, most of the long segments during the warmer months follow the exponential decay model, whereas more segments during the winter months follow the mean model. Some of the short segments during the summer months were assigned a mean model, which can be explained by the fact that these short segments, though displaying a decreasing pattern, do not fit in the typical shape of an exponential decay curve. For a more intuitive description, the estimated decay parameter $\theta$ is first transformed into the decay rate via $\gamma = \exp(-\exp(\theta))$, which takes a value between 0 and 1, and then converted into the e-folding decay time scale $\omega$. Among the fitted exponential decay models, the median decay rate for HOAL 01 is 0.993, with the first and third quartiles being 0.986 and 0.996. This corresponds to a median decay time scale of 12.036 days, and quartiles of 6.022 and 23.280 days. The median decay rate for HOAL 05 is 0.990, with the first and third quartiles being 0.983 and 0.995, which corresponds to a median decay time scale of 8.824 days, and quartiles of 4.818 days and 17.951 days. The estimated median e-folding decay time scales are consistent with the scales shown in the global map of \cite{SoilDrydown}.  

\subsection{NEON soil moisture time series}

The National Ecological Observatory Network (NEON) in the United States of America is a continental-scale observation facility operated by Battelle and designed to collect long-term ecological data \cite{NEONsoil}. The soil moisture time series data from NEON are available publicly and can be accessed from \url{https://data.neonscience.org/data-products/DP1.00094.001}. A brief introduction to the field sites is given below. 

The Central Plains Experimental Range (CPER) is a terrestrial NEON field site located in Weld County on the Western boundary of the Pawnee National Grasslands in Colorado. The region is known for dry, hot summers and cold winters. The combination of high elevation and mountain ranges is what mainly drives this region’s climate. the mean annual temperature is $8.6^{\circ}\mathrm{C} \, (47.5^{\circ}\mathrm{F})$ and the mean annual precipitation is 344 mm \citep{CPER}. The Talladega National Forest (TALL) is a terrestrial NEON field site located within the Oakmulgee District of the Talladega National Forest in west-central Alabama, where it has a subtropical climate with hot summers and mild winters. The warm, moist air contributes to the formation of convection storms and thunderstorms, causing major precipitation pulses and flooding. The average annual temperature is $17.2^{\circ}\mathrm{C}  \, (63^{\circ}\mathrm{F}$) and the average annual precipitation is 1380 mm \citep{ONAQ}.

\begin{figure}[!htb]
\begin{center}
\includegraphics[width=5.6in]{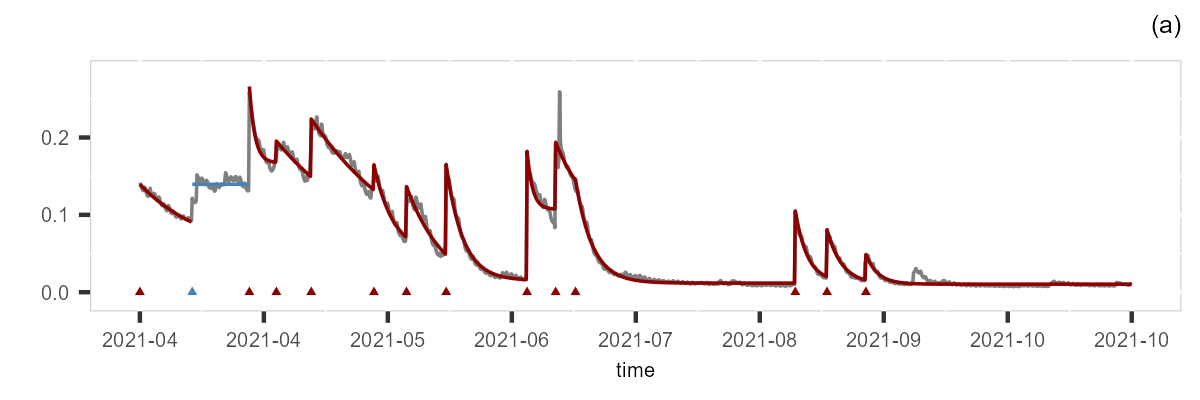}
\includegraphics[width=5.6in]{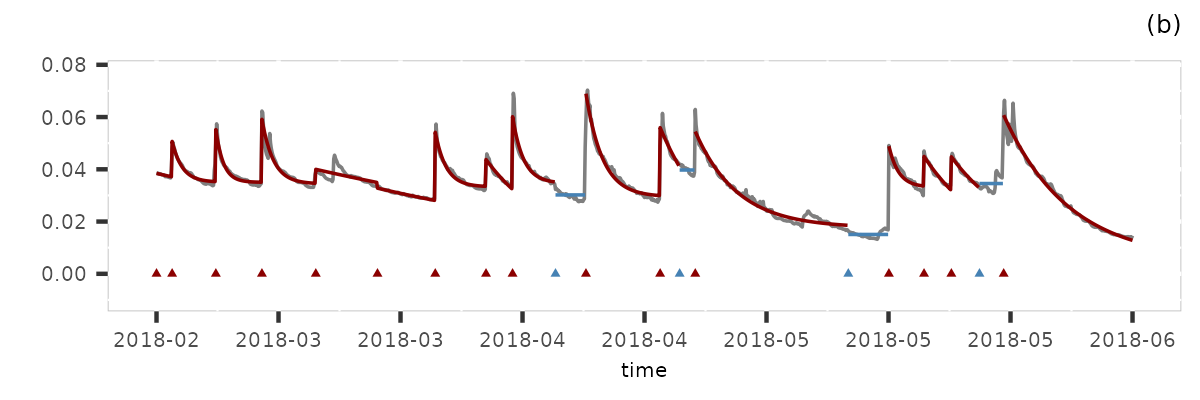}
\captionsetup{labelfont=bf, font=small}
\caption{The soil moisture time series from (a) NEON site CPER and (b) NEON site TALL, with the estimated changepoints and the fitted segments. The colour red represents the exponential decay model, and the colour blue represents the mean model.}
\label{fig:ApplicationNEON}
\end{center}
\end{figure}

As shown in Figure \ref{fig:ApplicationNEON}, the soil moisture time series from TALL is dominated by frequent peaks and drydown events, whereas the time series from CPER displays more versatile patterns. Similar to the analysis of the HOAL data, two candidate models, (a) an exponential decay model and (b) a mean model, are used, and the two models are assumed to have equal probability. The possibility of assigning a mean model gives flexibility when the exponential decay assumption becomes too unrealistic for some segments, despite the dominance of the decay patterns. A geometric run-length distribution was used, with the probability of a changepoint being 0.01. Prior distributions of the model parameters were selected based on the characteristics of the data; parameters in the inverse Gamma prior distribution were selected to reflect the scale of the noises in the time series. The results are shown in Figure \ref{fig:ApplicationNEON}. 

In general, the resulting models fit the time series well. There are missed changepoints and potentially misspecified segments, but the models captured the typical drydown segments closely, and the parameter estimation appears to be reasonable. The estimated median e-folding decay time scale in site CPER is 3.51 days. The first and third quartiles of the decay time scale are 2.56 and 8.49 days, respectively. The estimated median e-folding decay time scale in site TALL is 5.37 days, and the first and third quartiles of 0.81 and 7.18 days. As with all real-world data, it can be difficult to say for certain whether the segments are assigned the correct models. However, the segments assigned with an exponential decay model display clear drydown patterns. 

In both applications, we were able to identify the segments that display clear exponential decay patterns and retrieve the soil moisture drydown parameters that are of main interest to the investigation of soil health. No additional information, such as precipitation and temperature time series, was required when implement the methods. It automatically distinguishes the segments that do not have a clear drydown patterns from the ones that satisfy the exponential decay assumption. This can be a major advantage as compared to the conventional drydown modelling, where (high quality) precipitation data is often needed, and the manual identification of drydown segments is inevitable \citep{SoilDrydown, SDdiff, SDglobal}. The proposed method is overall more flexible than conventional method in modelling soil moisture time series with complex patterns.

\section{Discussion} \label{sec:Discussion}

This manuscript proposed two extensions of the Bayesian online changepoint detection algorithm to model time series that show complex patterns. This was motivated by the application problem in soil moisture drydown analysis, where distinctively different patterns are found across segments and the models used to describe these patterns contain unknown parameters that cannot be marginalised easily. The extensions make use of particle filter and online gradient descent to estimate the unknown parameters and obtain the marginal log-likelihood. Both extensions make use of the sequential inference nature of the BOCPD algorithm and can be used in both online and offline settings. The resulting parameter estimation procedure provides not only interesting information on the application problem but also indications on whether the segmentation makes sense. This is helpful in real-world applications when the true locations of the changepoints are unknown, because an unreasonable parameter estimate can signal a problem in the segmentation and the need to improve modelling. 

Although the extensions are motivated by the particular problem of modelling soil moisture data, they are in fact generic and can be applied to different types of time series that display alternating patterns. Among the extensions, BOCPD with a particle filter requires more computational time, but can potentially achieve better parameter estimation with associated uncertainty measure. On the other hand, whilst it is harder to assess convergence of parameter estimation, BOCPD with online gradient descent can be computed on the fly and hence can be attractive to the modelling of long time series. The investigation using simulated data provided some indications on this matter, although this manuscript did not attempt to provide a theoretical understanding. In practice, BOCPD+PF is more suited to shorter time series and when parameter estimation is of greater interest; whereas BOCPD+OG is more suited to longer time series or when changepoints estimation is of primary concern. 

The application of the proposed method was centered around the soil moisture time series data from ground-based sensors in the fields. In particular, we identified the drydown segments from the soil moisture time series and associated drydown parameters. Two candidate models, including an exponential decay model and a mean model, were used to describe the patterns in the time series data. The first analysis of HOAL data using BOCPD+OG extension was able to encode the changing winter-summer patterns in the time series data and automatically estimated drydown time scales that match the literature. The second analysis of NEON data identified the drydown curves and provided parameter estimates with uncertainty, which can be used in further analysis. Due to the complexity of the data, some changepoints were missed, and some segments were possibly assigned the wrong model. However, the application indicated the potential of the proposed methods in exploring the changing patterns in a long time series and providing estimates of key parameters. 

A key advantage is that the method can be applied to streams of soil moisture time series data automatically, once the initial input is given, such as the priors and global parameters. As an online algorithm, it can keep generating useful information as more data is collected. This can be attractive for the real-time monitoring of soil data via sensors, satellites, etc. Through further analysis of the estimated parameters and changepoint locations over a longer period of time, it may be possible to identify long-term changes in soils or flag unusual events or measurement errors. This could be of great interest to soil science and beyond. 

Implementation of the proposed method involves the selection of global parameters and prior parameters, which may require some sensitivity analysis. Therefore, the use of the method by scientists from a non-statistical background can be challenging. One important task for the future is to make the selection more systematic, such as creating a hyper-parameter selection protocol. Further reducing the computational cost, especially in the extension with the particle filter, is another area that requires future improvement. The current application capped the maximum number of changepoints kept at each iteration to be larger than the expected number of changepoints in the time series. A smaller number may be plausible, and should reduce the computation time, but will require verification. Alternatively, developing a pruning method based on e.g., the penalised exact linear time \citep{PELT}, may help reducing the computational cost further. Some intuition of the pruning method is given in the supplemental document. Finally, in terms of the soil moisture drydown analysis, it may be worth exploring candidate models that can capture more flexible patterns.

\section*{Acknowledgements}

The authors gratefully acknowledge the support from the UKRI-funded project Signals in the Soil (Grant No. NE/T012307/1).


\end{document}


\maketitle

\appendix

\section{Additional results from the illustration on simulated data} \label{sec:IllustrationMore}

As mentioned in section 2.2 of the main manuscript, backward simulation procedure can be applied to visualise the approximate posterior distributions of the changepoint locations. For an illustration using the simulated data (S1, S2, S3 an S4), backward simulation was repeated 500 times to obtain 500 changepoint configurations based on the filtering distributions obtained from the BOCPD + PF and BOCPD + OG algorithms for each of the four scenarios for each of the four scenarios. The count of the simulated changepoints out of 500 configuration were plotted in the figures below. 

\begin{figure}[!htb]
\begin{minipage}{0.5\textwidth}
\centering
\includegraphics[width=3.2in]{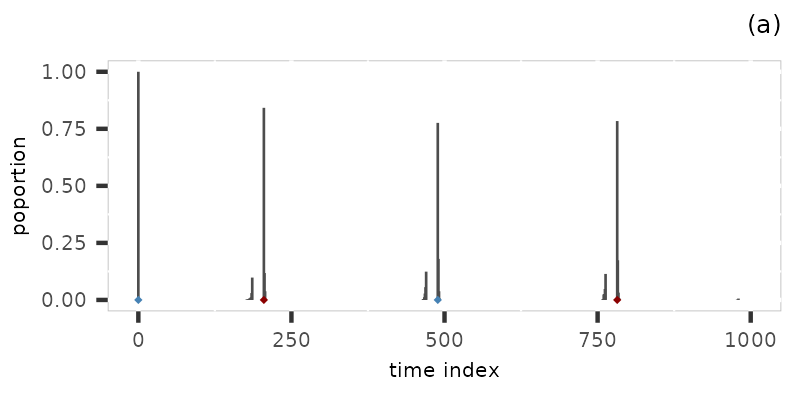}
\includegraphics[width=3.2in]{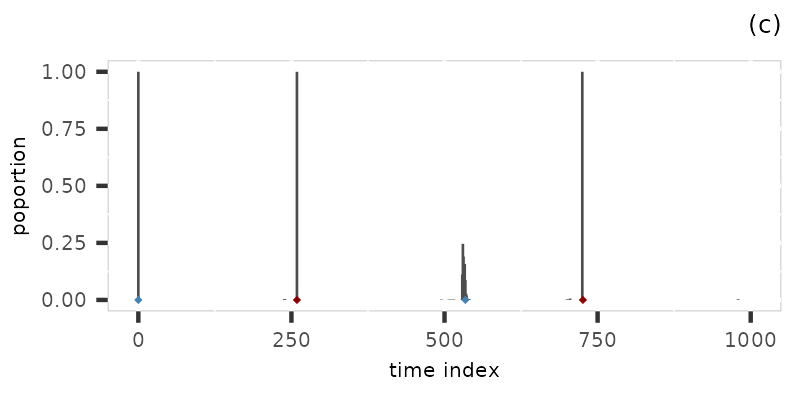}
\end{minipage} %
\begin{minipage}{0.5\textwidth}
\includegraphics[width=3.2in]{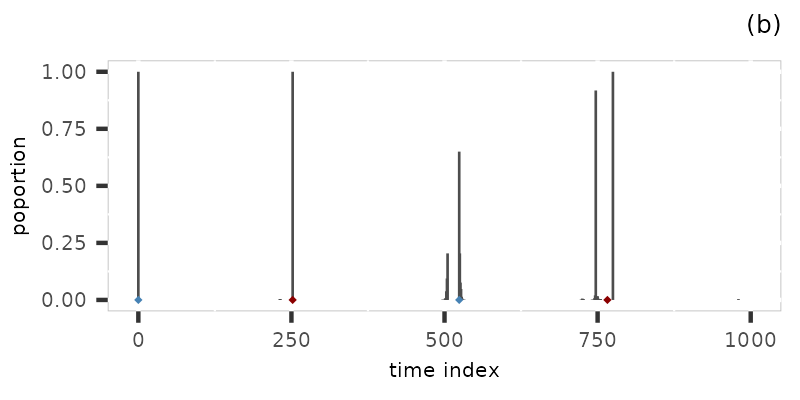}
\includegraphics[width=3.2in]{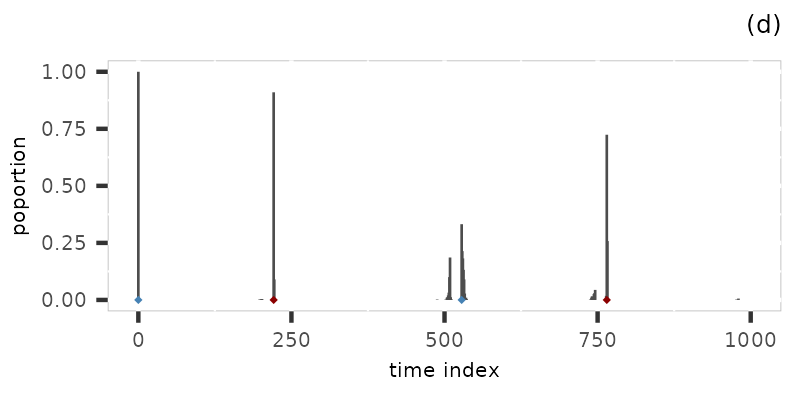}
\end{minipage} 
\captionsetup{labelfont=bf, font=small}
\caption{The proportion of each time point being identified as a changepoint from the 500 repetitions of the backward simulation using the result from BPCPD+PF, for scenarios (a) S1, (b) S2, (c) S3 and (4) S4. }
\label{fig:ScenariosPostPF1}
\end{figure}

\begin{figure}[!htb]
\begin{minipage}{0.5\textwidth}
\centering
\includegraphics[width=3.2in]{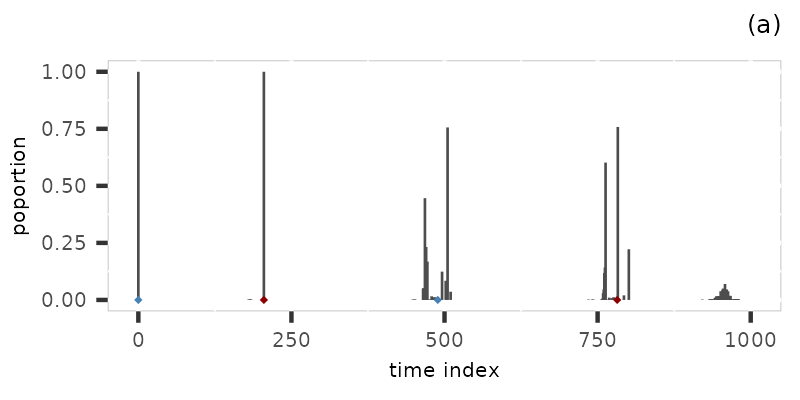}
\includegraphics[width=3.2in]{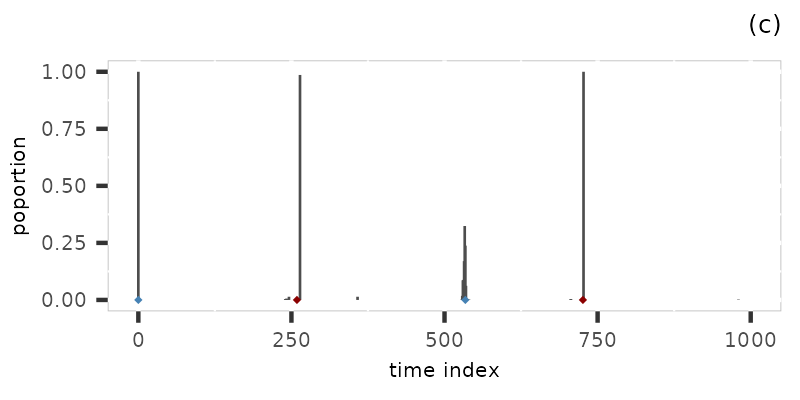}
\end{minipage} %
\begin{minipage}{0.5\textwidth}
\includegraphics[width=3.2in]{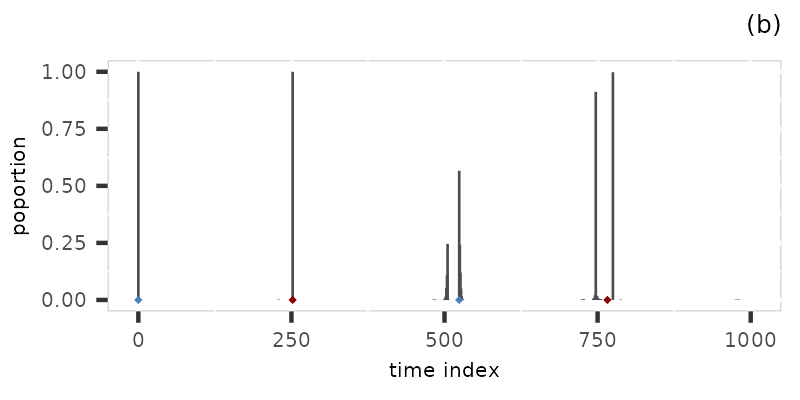}
\includegraphics[width=3.2in]{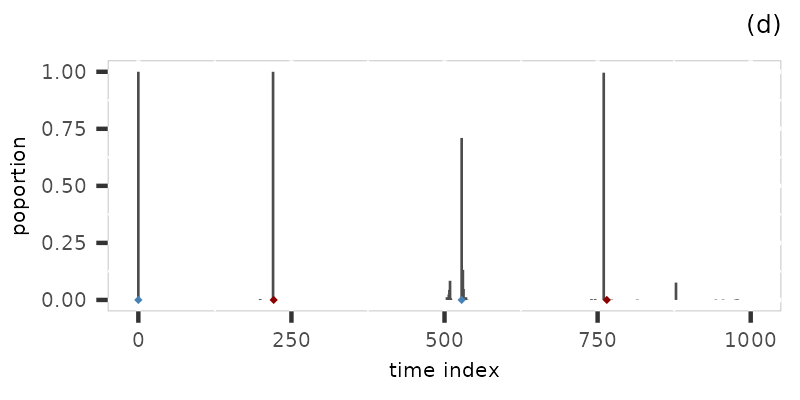}
\end{minipage} 
\captionsetup{labelfont=bf, font=small}
\caption{The proportion of each time point being identified as a changepoint from the 500 repetitions of the backward simulation using the result from BPCPD+OG, for scenarios (a) S1, (b) S2, (c) S3 and (4) S4. }
\label{fig:ScenariosPostPF2}
\end{figure}

\section{Using truncated Normal priors for the model coefficients} \label{sec:TruncNorm}

To avoid unrealistic estimation of the linear coefficients, one way is to use the truncated Normal distributions as priors. However, this will change the typical Normal - inverse Gamma priors combination for linear regression models, and will make the integration of marginal likelihood challenging. To see this, consider the marginal likelihood of the drydown model given the decay parameter $\theta_{s}$ (whether as a particle during the SMC or as an estimation during the online gradient descent), 
\begin{equation}
    \int_{\Omega_{\sigma^{2}}} \int_{\Omega_{\beta}} f(\bm{y}_{(s+1):t} \,| \, \theta_{s}, \bm{\beta}, \sigma^{2}) \pi(\bm{\beta} \,| \, \sigma^{2}) \pi (\sigma^{2}) d\bm{\beta} d \sigma^{2} \; .
    \label{eqn:Normal-invGamma}
\end{equation}
In a typical setting, for $\bm{\beta} = (\beta_{0}, \beta_{1})^{\top}$, $\Omega_{\beta} = \mathbb{R}^{2}$. The priors would be $\bm{\beta} \,| \, \sigma^{2} \sim \mathcal{N}(\mu_{0}, \sigma^{2} V_{0}) $, $\sigma^{2} \sim \mathrm{invGamma}(u, v)$, and the integration (\ref{eqn:Normal-invGamma}) can be calculated in closed form. This is no longer the case if the truncated Normal priors are used. Let $\phi(\beta |  \mu_{0}, \sigma^{2} V_{0})$ and $\Phi(\beta \in \Omega_{\beta} |  \mu_{0}, \sigma^{2} V_{0})$ be the density function and cumulative distribution function of Normal distribution. Then $\pi(\bm{\beta} \,| \, \sigma^{2}) = \phi(\beta |  \mu_{0}, \sigma^{2} V_{0}) / \Phi(\beta \in \Omega_{\beta} |  \mu_{0}, \sigma^{2} V_{0})$, and $\pi(\bm{\beta} \,| \, \sigma^{2}, \bm{y}_{(s+1):t}) = \phi(\beta |  \mu_{1}, \sigma^{2} V_{1}) / \Phi(\beta \in \Omega_{\beta} |  \mu_{1}, \sigma^{2} V_{1})$, where $\mu_{1}$ and $\sigma^{2} V_{1}$ are posterior mean and variance. The integration (\ref{eqn:Normal-invGamma}) becomes 
\begin{equation*}
    \int_{\Omega_{\sigma^{2}}} \int_{\Omega_{\beta}} \left( \frac{1}{\sqrt{2 \pi \sigma^{2}}} \right)^{2} \frac{1}{\sqrt{|V_{1}|}} \exp \left\{ - \frac{1}{2\sigma^{2}} (\bm{\beta} - \mu_{1})^{\top} V_{1}^{-1} (\bm{\beta} - \mu_{1}) \right \} d\bm{\beta} \, F(\sigma^{2}) \pi (\sigma^{2}) d \sigma^{2} \; ,
    \label{eqn:TruncNormal}
\end{equation*}
where $F(\sigma^{2}) = \left( \frac{1}{\sqrt{2 \pi \sigma^{2}}} \right)^{t-s} \sqrt{\frac{|V_{1}|}{|V_{0}|}} \, \frac{1}{\Phi(\beta \in \Omega_{\beta} |  \mu_{0}, \sigma^{2} V_{0})}$. The inner integration with respect to $\bm{\beta}$ gives $\Phi(\beta \in \Omega_{\beta} |  \mu_{1}, \sigma^{2} V_{1})$. As a result, the outer integration with respect to $\sigma^{2}$ involves the ratio of two Normal cumulative distribution functions, $\Phi(\beta \in \Omega_{\beta} |  \mu_{1}, \sigma^{2} V_{1}) / \Phi(\beta \in \Omega_{\beta} |  \mu_{0}, \sigma^{2} V_{0})$, which makes the analytical solution to the integration impossible. To avoid further complication to the application problem, this study chooses to approximate this ratio using an empirical estimation of $\sigma^{2}$, and treat it as a quantity that does not involve the unknown $\sigma^{2}$. The reasons are two folds. First of all, it is plausible to get the empirical estimation. Secondly, even though the ratio is not exact, it can still provide some adjustments to the likelihood to reflect the impact of the truncated distribution.

\section{Possibility of using pruning to reduce computational cost}

The pruning strategy, known as the penalised exact linear time (PELT) algorithm, introduced in \cite{PELT}, is a popular method of reducing the computational cost in changepoint detection problems solved via dynamic programming. It provides an exact solution to the problem, as opposed to the approximate solution by the resampling method. Motivated by the similarity between the forward filtering recursion of the BOCPD algorithm and the forward recursion of dynamic programming, the possibility of using a PELT-based method to reduce the computational cost of BOCPD is investigated. The main question is to find a suitable cost function that satisfies the condition $C(\bm{y}_{(s+1):r}) + C(\bm{y}_{(r+1):t}) + K \leq C(\bm{y}_{(s+1):t})$ for some constant $K$. That is, adding a changepoint $r$ to a segment from $s+1$ to $t$ will not increase the overall cost of the segment, given the constant $K$. 

The PELT method was initially developed in the context of maximum log-likelihood estimation in \cite{PELT}, where $K=0$. For the BOCPD, a potential choice of the cost function is the negative log marginal likelihood, minus the log run-length distribution, $C(\bm{y}_{(s+1):t}) = - \log [L(s, t, m)] - \log [g(t-s)]$. This is motivated by the recursive calculation in the maximum a posteriori estimation. However, the constant $K$ is no longer $0$ in this case, but a quantity related to both the log marginal likelihood and the log run-length distribution. When a geometric distribution is used as the run-length distribution, there is  $\log [g(t-r)] + \log [g(r-s)] = \log [g(t-s)] + \log ( \frac{\eta}{1-\eta} )$, where $\eta$ is the probability of a change, so the part of the constant related to the run-length is $-\log ( \frac{\eta}{1-\eta} )$. For a log marginal likelihood obtained by integrating over the unknown parameter $\theta$ given a prior $\pi(\theta)$, the quantity that ensures that an arbitrary split of the segment will not increase the negative overall log-likelihood is related to the KL divergence between the posterior and the prior of $\theta$ given the information of the segment $\bm{y}_{(s+1):r}$. Let $C^{\ast}(\bm{y}_{(s+1):t}) = - \log \left[ f(\bm{y}_{(s+1):t}) \right]$, it can be shown that
\begin{equation*}
\label{eqn:marginal_cost}
    C^{\ast}(\bm{y}_{(s+1):r}) + C^{\ast}(\bm{y}_{(r+1):t}) - \text{KL}(p(\theta | \bm{y}_{(s+1):t}) \, || \, \pi(\theta)) \leq C^{\ast}(\bm{y}_{(s+1):t}) \, ,
\end{equation*}
where the KL divergence is
\begin{equation*}
    \text{KL}(p(\theta | \bm{y}_{(s+1):t}) \, || \, \pi(\theta)) = \int \log \left[ \frac{p(\theta | \bm{y}_{(s+1):t})}{\pi(\theta)} \right] \, p(\theta | \bm{y}_{(s+1):t}) \, d \theta \, .
\end{equation*}
Therefore, in order to use the pruning strategy in practice, one needs to find an upper bound on the KL divergence $\text{KL}(p(\theta | \bm{y}_{(s+1):t}) \, || \, \pi(\theta))$ for any segment $\bm{y}_{(s+1):t}$. Under some assumptions, such as (i) the likelihood function is continuous in $\theta$, (ii) the expectation $\mathbf{E}_{p(\theta)}\left[ L(\theta; \bm{y}_{(s+1):t}) \right] $ exists and is finite, and (iii) the marginalised likelihood is log-concave, so that $C^{\ast}(\bm{y}_{(s+1):t}) = -\log [ L(\theta; \bm{y}_{(s+1):t}) ]$ is convex, it can be shown using properties of Lipschitz continuous functions that the KL divergence is indeed upper bounded. However, finding such an upper bound in practice is extremely challenging.